\documentclass[twocolumn,aps,prl,tightenlines]{revtex4}
\usepackage[latin9]{inputenc}
\usepackage{amsmath}
\usepackage{amssymb}
\usepackage{graphicx}
\usepackage{esint}

\makeatletter
\@ifundefined{textcolor}{}
{%
 \definecolor{BLACK}{gray}{0}
 \definecolor{WHITE}{gray}{1}
 \definecolor{RED}{rgb}{1,0,0}
 \definecolor{GREEN}{rgb}{0,1,0}
 \definecolor{BLUE}{rgb}{0,0,1}
 \definecolor{CYAN}{cmyk}{1,0,0,0}
 \definecolor{MAGENTA}{cmyk}{0,1,0,0}
 \definecolor{YELLOW}{cmyk}{0,0,1,0}
}

\usepackage{mathrsfs}
\usepackage{bbm}
\usepackage{bbold}
\usepackage{subfigure}
\newcommand{\rmd}{{\rm d}}

\newcommand{\fvec}[1]{\boldsymbol{#1}}

\newcommand{\half}{\frac{1}{2}}

\makeatother

\begin{document}

\title{Superconductivity in FeSe thin films driven by the interplay between
nematic fluctuations and spin-orbit coupling}

\author{Jian Kang}
\email{jkang@umn.edu}
\affiliation{School of Physics and Astronomy, University of Minnesota, Minneapolis,
MN 55455, USA}

\author{Rafael M. Fernandes}

\affiliation{School of Physics and Astronomy, University of Minnesota, Minneapolis,
MN 55455, USA}
\begin{abstract}
The origin of the high-temperature superconducting state observed
in FeSe thin films, whose phase diagram displays no sign of magnetic
order, remains a hotly debated topic. Here we investigate whether
fluctuations arising due to the proximity to a nematic phase, which
is observed in the phase diagram of this material, can promote superconductivity.
We find that nematic fluctuations alone promote a highly degenerate
pairing state, in which both $s$-wave and $d$-wave symmetries are
equally favored, and $T_{c}$ is consequently suppressed. However,
the presence of a sizable spin-orbit coupling or inversion symmetry-breaking
at the film interface lifts this harmful degeneracy and selects the
$s$-wave state, in agreement with recent experimental proposals.
The resulting gap function displays a weak anisotropy, which agrees
with experiments in monolayer FeSe and intercalated Li$_{1-x}$(OH)$_{x}$FeSe.
\end{abstract}
\maketitle

In most iron-based superconductors (FeSC), superconductivity is found
in close proximity to a magnetically ordered state, suggesting that
magnetic fluctuations play an important role in binding the Cooper
pairs \cite{Mazin08,Hirschfeld11,Chubukov11,Chubukov_review}. Indeed,
the fact that the Fermi surface of these materials is composed of
small hole pockets and electron pockets separated by the magnetic
ordering vector led to the proposal of a sign-changing $s^{+-}$ wave
state, in which the gap function has different signs in the hole and
in the electron pockets. However, the recent observation of superconductivity
over $70$ K in monolayer FeSe brought new challenges to the field
\cite{Xue12,Feng12,Zhou13,Feng13,Xue14,Jia14,Wang14,Ding15,FengNC14,Chen16}.
In contrast to the standard FeSC, no long-range magnetic order is
observed in thin films or even bulk FeSe \cite{Khasanov09}, and the
Fermi surface of monolayer FeSe consists of electron pockets only
\cite{Zhou13,Feng12,Shen14,Ding15}. Since $T_{c}$ in monolayer FeSe
is the highest among all FeSC, the elucidation of its origin is a
fundamental step in the search for higher $T_{c}$ in these systems.

One of the proposed scenarios to explain the dramatic ten-fold increase
of $T_{c}$ in monolayer FeSe with respect to the $8$ K value in
bulk FeSe~\cite{HSU08} was the strong coupling to an optical phonon
mode of the SrTiO$_{3}$ (STO) substrate~\cite{Shen14,Lee12,Zhao16},
which is manifested by replica bands observed in ARPES~\cite{Shen14}.
Although such a coupling can certainly enhance $T_{c}$~\cite{Rademaker16,DHLee_STO,Millis16,Johnston16,Dolgov16},
recent experiments indicate that the STO substrate may not be essential
to achieve the high-$T_{c}$ state. In particular, $T_{c}$ up to
$40$ K was observed in electrostatically-gated films of FeSe with
different thickness grown both on STO and MgO substrates \cite{Tsukazaki16}.
Similar values of $T_{c}$ were reported in FeSe coated with potassium
\cite{Takahashi15,ShenSC15} and in the bulk sample Li$_{1-x}$(OH)$_{x}$FeSe~\cite{FengPRB15,Zhou16},
which consists of intercalated FeSe layers. In common to all these
systems is the fact that their Fermi surface consists of electron
pockets only, suggesting that doping by negative charge carriers plays
a fundamental role in stabilizing the high-$T_{c}$ state.

Importantly, recent experiments in K-coated bulk FeSe \cite{ShenSC15}
revealed that, besides shifting the chemical potential, electron-doping
also suppresses the nematic order observed in undoped bulk FeSe at
$T_{\mathrm{nem}}\approx90$ K~\cite{Coldea15}. In the nematic state,
whose origin remains hotly debated~\cite{RMF15,RMF16,Si15,DHLee_FeSe15,Glasbrenner15},
the $x$ and $y$ in-plane directions become inequivalent and orbital
order emerges. Remarkably, the highest $T_{c}$ in the phase diagram
of K-coated FeSe is observed near the region where $T_{\mathrm{nem}}$
nearly vanishes. Similarly, in the case of FeSe thin films grown on
STO, nematic order is observed over a wide range of film thickness~\cite{ShenNem15,Feng16},
but not in the monolayer case~\cite{Hoffman16}. These observations,
combined with the absence of magnetic order in these systems, begs
the question of whether nematic fluctuations can provide a sensible
mechanism to explain the superconductivity of thin films of FeSe \cite{ShenSC15,Yamase13,DHLee_STO,Vishwanath15}.

In this paper, we show that nematic fluctuations alone favor degenerate
$s$-wave ($A_{1g}$) and $d$-wave ($B_{2g}$) superconducting states
in FeSe thin films. This degeneracy stems from the fact that while
the two electron pockets are separated by the momentum $\fvec Q_{M}=\left(\pi,\pi\right)$,
nematic fluctuations are peaked at $\fvec Q_{\mathrm{nem}}=0$. More
importantly, the SC ground state manifold has an enlarged $U(1)\times U(1)$
degeneracy, which is very detrimental to SC, since fluctuations of
one SC channel strongly suppress long-range order in the other SC
channel. Remarkably, this degeneracy is removed by the sizable spin-orbit
coupling (SOC) observed in these compounds \cite{Zhigadlo16}, which
lift the pairing frustration and selects $s$-wave over $d$-wave,
stabilizing a SC state at higher temperatures. In thin films, the
inversion symmetry-breaking (ISB) at the interface also contributes
significantly to this degeneracy lifting. Interestingly, recent experiments
propose that an $s$-wave state is realized in FeSe thin films \cite{FengNP15}.
We also find that, when the SOC and/or ISB energy scales are larger
than the energy scale associated with the mismatch between the two
electron pockets, a nearly isotropic gap appears at the electron pockets,
whose angular dependence agrees with ARPES and STM measurements in
FeSe thin films~\cite{ShenSC15,FengPRL14} and intercalated Li$_{1-x}$(OH)$_{x}$FeSe~\cite{Wen16}.

\begin{figure}[htbp]
\includegraphics[width=0.99\columnwidth]{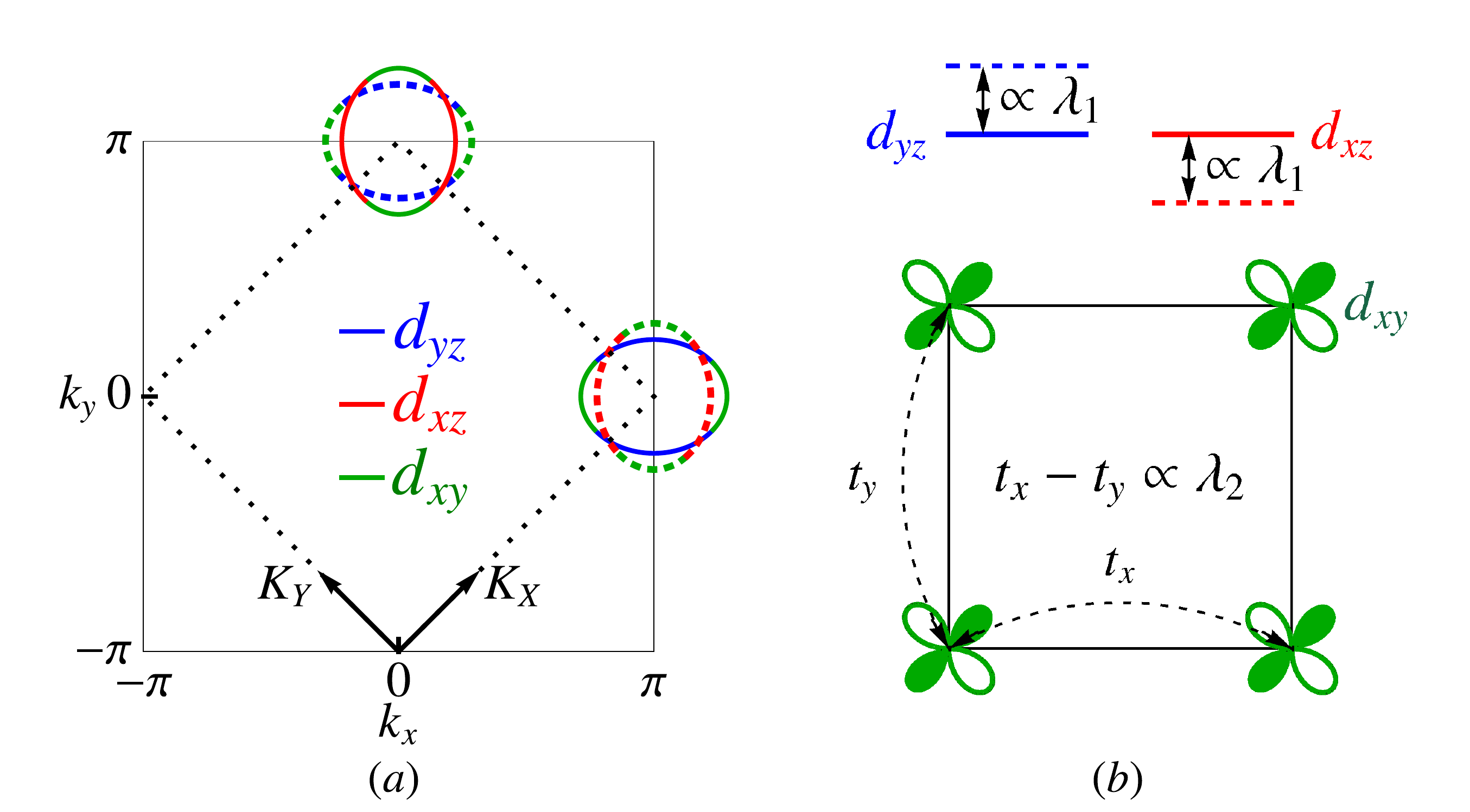} \protect\protect\caption{(a) Fermi surface (FS) of a thin film of FeSe, consisting only of
electron pockets, in the unfolded (solid lines) and folded (dotted
lines) Brillouin zones. The color around the FS indicates the orbital
that contributes the largest spectral weight. (b) The two different
nematic couplings: $\lambda_{1}$ couples to the on-site energy difference
between the $d_{xz}$ and $d_{yz}$ orbitals, whereas $\lambda_{2}$
couples to the anisotropic hopping between nearest-neighbor $d_{xy}$
orbitals. \label{Fig:IronFS}}
\end{figure}

\emph{Microscopic model}\ \ \ We start with the full five-orbital
tight-binding model in the 1-Fe Brillouin zone and project it on the
subspace of the $d_{xz}$, $d_{yz}$, and $d_{xy}$ orbitals, which
give the largest contribution to the Fermi surface. In particular,
while the $X$ electron pocket centered at $\fvec Q_{X}=(\pi,0)$
has $d_{yz}/d_{xy}$ orbital content, the $Y$ electron pocket centered
at $\fvec Q_{Y}=(0,\pi)$ has $d_{xz}/d_{xy}$ content (see Fig.~\ref{Fig:IronFS}a).
Following Ref.~\cite{Oskar13}, we expand the projected tight-binding
matrix in powers of the momentum measured relative to $\fvec Q_{X}$
and $\fvec Q_{Y}$. Defining two spinors corresponding to each electron
pocket:
\begin{eqnarray}
\Psi_{X}(\fvec k) & \approx & \left(d_{yz}(\fvec k+\fvec Q_{X})\ ,\ d_{xy}(\fvec k+\fvec Q_{X})\right)^{T}\nonumber \\
\Psi_{Y}(\fvec k) & \approx & \left(d_{xz}(\fvec k+\fvec Q_{Y})\ ,\ d_{xy}(\fvec k+\fvec Q_{Y})\right)^{T}\label{spinors}
\end{eqnarray}
the non-interacting Hamiltonian is written as $\mathcal{H}_{0}=\sum\limits _{\fvec k,i=X,Y}\Psi_{i}^{\dagger}(\fvec k)\hat{H}_{i}(\fvec k)\Psi_{i}(\fvec k)$
where $\hat{H}_{i}$ are $2\times2$ matrices in spinor space (see
the supplementary material SM). The $B_{2g}$ nematic order parameter
is described by the bosonic field $\phi_{q}$, with $q=\left(\Omega_{n},\fvec q\right)$,
whereas the nematic fluctuations are given by the nematic susceptibility
$\chi_{\mathrm{nem}}(\fvec q,\Omega_{n})$. For our analysis, it is
not necessary to specify the origin of the nematic order parameter,
but rather how it couples to the electronic states. As discussed in
Ref.~\cite{Vafek14}, there are two possible nematic couplings: $\lambda_{1}$,
which couples $\phi_{q}$ to the onsite energy difference between
the $d_{xz}$ and $d_{yz}$ orbitals, and $\lambda_{2}$, which couples
$\phi_{q}$ to the hopping anisotropy between nearest-neighbor $d_{xy}$
orbitals (see Fig.~\ref{Fig:IronFS}b):

\begin{equation}
\mathcal{H}_{\mathrm{int}}=\sum_{\fvec q,i=X,Y}\phi_{-\mathbf{q}}\Psi_{i}^{\dagger}(\fvec k)\hat{\lambda}_{i}^{\mathrm{nem}}\Psi_{i}(\fvec k+\fvec q)\label{H_int}
\end{equation}
with $\hat{\lambda}_{i}^{\mathrm{nem}}=\pm\mathrm{diag}\left(\lambda_{1},\lambda_{2}\right)$,
where the plus (minus) sign refers to $i=X$ ($i=Y)$. Here, we focus
on the effect of short-ranged frequency independent nematic fluctuations
and approximate $\chi_{\mathrm{nem}}(\fvec q,\Omega_{n})$ by its
zero momentum and zero frequency value. The first approximation is
justified due to the smallness of the electron pockets, whereas the
second one is reasonable as long as the system is not too close to
a nematic quantum critical point~\cite{Schattner15,DHLee15,Vishwanath15}.
Note that renormalization-group calculations on a related microscopic
model support the idea that the disappearance of the central hole
pockets suppresses nematic order \cite{RMF16}.

\emph{Superconducting instability}\ \ \ We decompose the pairing
states in terms of the different irreducible representations of the
space group of the FeSe plane, $P4/nmm$ (see Ref.~\cite{Oskar13}
and the SM), and focus on the two leading pairing channels, which
belong to the singlet $s$-wave ($A_{1g}$) and $d$-wave ($B_{2g}$)
symmetry representations \cite{note}:
\begin{equation}
\Psi_{X}^{T}\begin{pmatrix}\Delta_{1} & 0\\
0 & \Delta_{2}
\end{pmatrix}\otimes i\sigma_{2}\Psi_{X}\pm\Psi_{Y}^{T}\begin{pmatrix}\Delta_{1} & 0\\
0 & \Delta_{2}
\end{pmatrix}\otimes i\sigma_{2}\Psi_{Y}\label{HSC}
\end{equation}
where the plus (minus) sign refers to $s$-wave ($d$-wave) pairing.
The gaps $\Delta_{1}$ and $\Delta_{2}$ correspond to intra-orbital
pairing within the $d_{xz}/d_{yz}$ orbitals and $d_{xy}$ orbitals,
respectively.$\Delta_{1}$ and $\Delta_{2}$ are found via the gap
equations:

\begin{equation}
\eta\hat{M}=\chi_{\mathrm{nem}}T\sum_{n,\fvec k}\left(\hat{\lambda}_{i}^{\mathrm{nem}}\right)^{T}\hat{G}_{-k,i}^{T}\hat{M}\hat{G}_{k,i}\hat{\lambda}_{i}^{\mathrm{nem}}\label{Delta_SC}
\end{equation}
where $\eta$ is the SC eigenvalue, $\hat{M}=\begin{pmatrix}\Delta_{1} & 0\\
0 & \Delta_{2}
\end{pmatrix}$, and $\hat{G}_{p,i}^{-1}=i\omega_{n}-\hat{H}_{i}\left(\mathbf{p}\right)$.
The SC transition temperature is obtained when $\eta=1$. Hereafter,
we set the value of $\left(\lambda_{1}^{2}+\lambda_{2}^{2}\right)\chi_{\mathrm{nem}}$
to yield $T_{c}=5$meV when $\lambda_{2}=0$.

\begin{figure}[htbp]
\centering \includegraphics[scale=0.4]{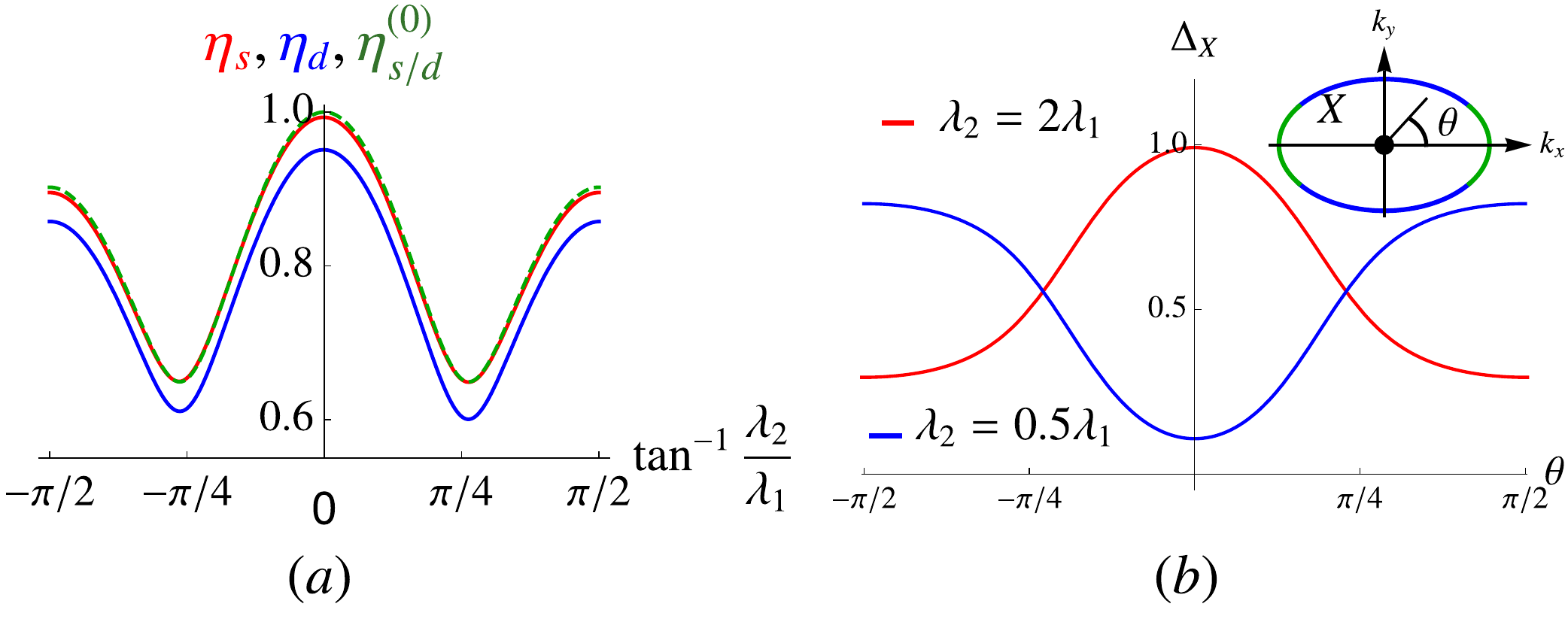} \protect\protect\caption{(a) The eigenvalue $\eta$ of the gap equation (\ref{Delta_SC}) as
function of the ratio between the two nematic couplings $\lambda_{2}/\lambda_{1}$.
Without SOC or ISB, the $s$-wave and $d$-wave solutions have the
same eigenvalue (dashed green curve, $\eta_{s/d}^{(0)}$ ). The presence
of SOC or ISB removes this degeneracy, making $s$-wave (red curve,
$\eta_{s}$) the leading pairing instability and $d$-wave (blue curve,
$\eta_{d}$) the subleading one. (b) Normalized SC gap along the $X$
electron pocket as function of the angle $\theta$ for different values
of $\lambda_{2}/\lambda_{1}$. \label{Fig:NemSC}}
\end{figure}

Solution of the gap equations reveals that for all ratios of the nematic
coupling constants $\lambda_{1}$ and $\lambda_{2}$, the superconducting
instabilities in the $s$-wave and $d$-wave channels are always degenerate,
as shown in Fig.~\ref{Fig:NemSC}a. Although the intra-orbital gaps
$\Delta_{1}$ and $\Delta_{2}$ are isotropic, the gaps projected
onto the Fermi pockets, $\Delta_{X}$ and $\Delta_{Y}$, acquire an
angle-dependence due to the orbital content of the Fermi pockets.
To illustrate this behavior, Fig.~\ref{Fig:NemSC}b shows $\Delta_{X}$
as function of the polar angle $\theta$. When $\lambda_{1}>\lambda_{2}$,
nematic fluctuations couple mainly to the $d_{xz}/d_{yz}$ orbitals;
as a result, $\Delta_{X}$ is proportional to the spectral weight
of the $d_{xz}/d_{yz}$ orbital on the $X$ pockets, which is maximum
around $\theta=\pm\pi/2$ (see Fig.~\ref{Fig:IronFS}a). Consequently,
$\Delta_{X}$ reaches its maximum at $\theta=\pm\pi/2$ and its minimum
at $\theta=0,\pi$. Conversely, for $\lambda_{1}<\lambda_{2}$, the
gap is maximum at $\theta=0,\pi$, where the spectral weight of the
$d_{xy}$ orbital on the $X$ pocket is maximum. Recent ARPES experiments
in FeSe suggest that $\lambda_{1}$ and $\lambda_{2}$ are comparable~\cite{Borisenko}.

In terms of the averaged gaps $\Delta_{X}$ and $\Delta_{Y}$, the
$s$-wave and $d$-wave solutions correspond to $\Delta_{s}=\frac{1}{2}(\Delta_{X}+\Delta_{Y})$
and $\Delta_{d}=\frac{1}{2}(\Delta_{X}-\Delta_{Y})$. Using this notation,
the degeneracy between $s$ and $d$ can be understood as a consequence
of the fact that nematic fluctuations, peaked at $\fvec Q_{\mathrm{nem}}=0$,
do not couple the gaps at the $X$ and $Y$ pockets, since they are
displaced by the momentum $\mathbf{Q}_{M}=\mathbf{Q}_{X}+\mathbf{Q}_{Y}=(\pi,\pi)$.
This suggests an enlarged $U(1)\times U(1)$ degeneracy of the SC
ground state manifold, corresponding to two decoupled SC order parameters.
To investigate the robustness of this enlarged degeneracy, we went
beyond the linearized gap equations and computed the superconducting
free energy to quartic order in the gaps (see SM), obtaining:
\begin{equation}
F_{SC}=a\left(|\Delta_{X}|^{2}+|\Delta_{Y}|^{2}\right)+\frac{u}{2}\left(|\Delta_{X}|^{4}+|\Delta_{Y}|^{4}\right)\label{F_SC}
\end{equation}

This form confirms that $\Delta_{X}$ and $\Delta_{Y}$ remain decoupled
to higher orders in $F_{SC}$. The consequences of this enlarged $U(1)\times U(1)$
degeneracy are severe: going beyond the mean-field approximation of
Eq.~(\ref{Delta_SC}), fluctuations of one SC channel suppress long-range
order in the other channel, i.e. $T_{c,s}-T_{c,0}\propto-\left\langle \Delta_{d}^{2}\right\rangle $.
Such a pairing frustration is therefore detrimental to SC \cite{DHLee_13,Fernandes13,Brydon14,YXWang16},
suggesting that nematic fluctuations alone do not provide an optimal
SC pairing mechanism in this system. Interestingly, previous investigations
of SC induced by nematic fluctuations in different models also found
nearly-degenerate states~\cite{Yamase13,Kivelson15}.

\emph{Spin-orbit coupling (SOC) and Inversion symmetry-breaking (ISB)}\ \ \ The
analysis above neglected a key property of the crystal structure of
the FeSe plane: Because of the puckering of the Se atoms above and
below the Fe square lattice, the actual crystallographic unit cell
contains 2 Fe atoms. As a result, in the 2-Fe Brillouin zone (the
folded BZ), the momentum $\fvec Q_{M}=\left(\pi,\pi\right)$ becomes
$\mathbf{\tilde{Q}}=0$ (hereafter the tilde denotes a wave-vector
in the folded BZ). Thus, the two electron pockets become centered
at the same momentum $\tilde{\mathbf{Q}}=\left(\pi,\pi\right)$ and
overlap, as shown by the dashed lines in Fig.~\ref{Fig:IronFS}a.

\begin{figure}[htbp]
\includegraphics[width=0.9\columnwidth]{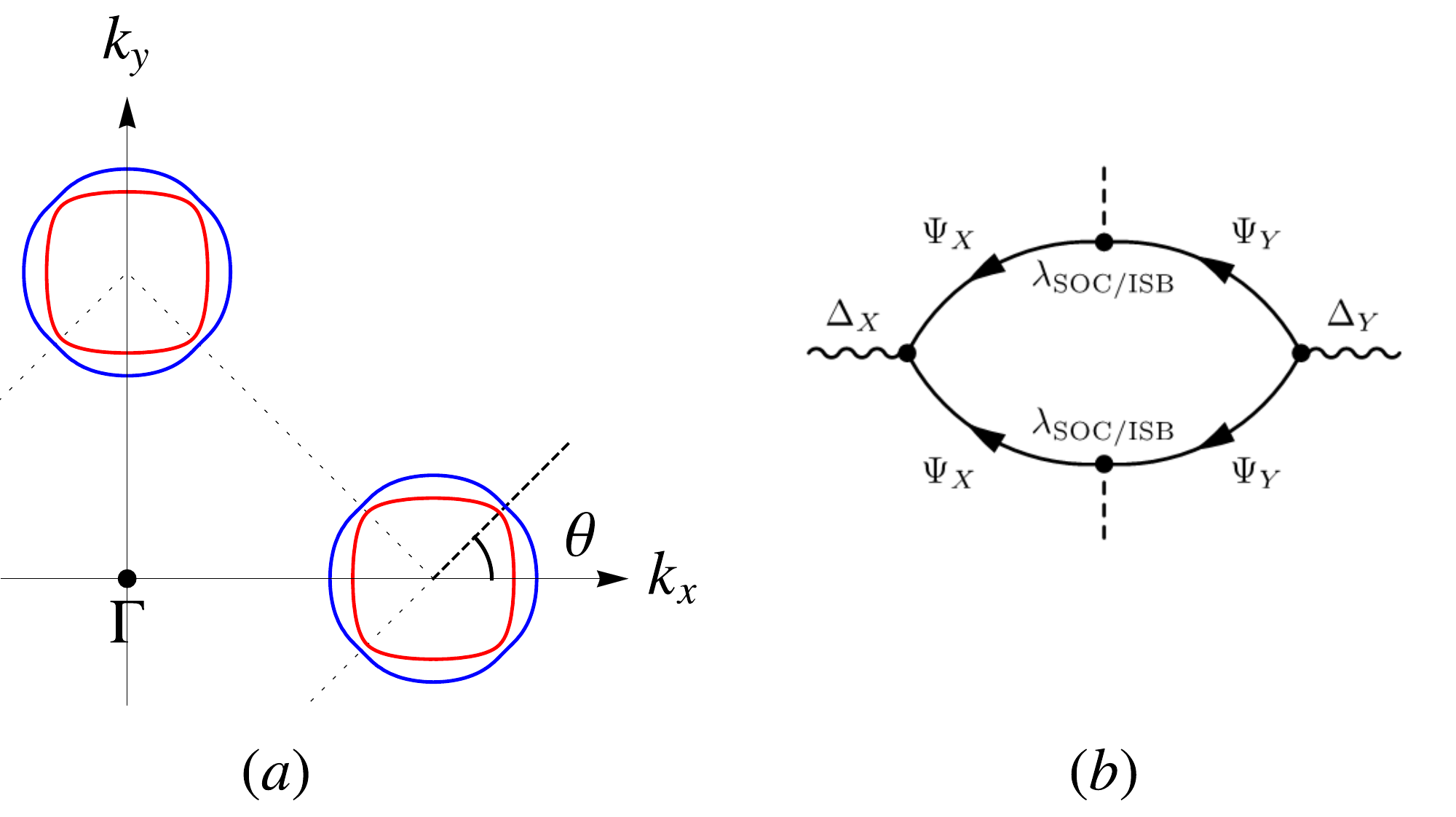} \protect\protect\caption{(a) The Fermi surface in the presence of SOC or ISB consists of split
inner (red) and outer (blue) electron pockets. (b) Feynman diagram
representing the coupling between the gaps in the two electron pockets
promoted by SOC or ISB. This coupling lifts the degeneracy between
$s$-wave and $d$-wave. \label{fig_SC_SOC}}
\end{figure}

This property opens up the possibility of coupling the $\Delta_{X}$
and $\Delta_{Y}$ gaps and removing the enlarged $U(1)\times U(1)$
degeneracy. At the non-interacting level, this is accomplished by
the atomic spin orbit coupling $\lambda_{\mathrm{SOC}}\fvec S\cdot\fvec L$,
which couples the $d_{xz}$ ($d_{yz}$) orbital associated with the
$Y$ ($X$) pocket to the $d_{xy}$ orbital associated with the $X$
($Y$) pocket \cite{Vafek14}:
\begin{equation}
\mathcal{H}_{\mathrm{SOC}}=\frac{i}{2}\lambda_{\mathrm{SOC}}\sum_{\fvec k}\Psi_{Y}^{\dag}\left(\tau_{+}\otimes\sigma_{1}+\tau_{-}\otimes\sigma_{2}\right)\Psi_{X}+h.c.\label{Eqn:HamSOC}
\end{equation}
where $\tau$ and $\sigma$ are Pauli matrices in spinor and spin
spaces, respectively. While in the normal state the SOC splits the
two overlapping elliptical electron pockets centered at $\tilde{\mathbf{Q}}=\left(\pi,\pi\right)$
into inner and outer pockets (see Fig.~\ref{fig_SC_SOC}a and the
ARPES data of \cite{Zhigadlo16}), in the SC state it couples the
gaps $\Delta_{X}$ and $\Delta_{Y}$. For $\lambda_{\mathrm{SOC}}$
small compared to $\epsilon_{m}$ -- the energy scale associated with
the mismatch between the $X$ and $Y$ electron pockets -- this coupling
is given perturbatively by the Feynman diagram of Fig.~\ref{fig_SC_SOC}b,
which gives the following contribution to the SC free energy of Eq.~(\ref{F_SC}):

\begin{equation}
\delta F_{SC}=\gamma\left(\Delta_{X}\Delta_{Y}^{*}+h.c\right)\label{delta_F_SC}
\end{equation}

As shown in the SM, $\gamma\propto-\lambda^{2}$, implying that the
SOC selects the $s$-wave state, with $\Delta_{X}$ and $\Delta_{Y}$
of the same sign, over the $d$-wave state, with $\Delta_{X}$ and
$\Delta_{Y}$ of opposite signs. More importantly, it lifts the $U(1)\times U(1)$
degeneracy between the two pairing states, suppressing the negative
interference of one pairing channel on the other. We confirmed this
general conclusion by evaluating explicitly the gap equations in the
$A_{1g}$ ($s$-wave) and $B_{2g}$ ($d$-wave) channels, finding
that $\eta_{s}>\eta_{d}$ for all values of the nematic coupling constants,
as shown in Fig.~\ref{Fig:NemSC}a. Note that the SOC induces triplet
components to these pairing states (see SM).

Having established that the $A_{1g}$ channel is the leading SC instability,
we now discuss the angular dependence of the gaps $\Delta_{i/o}$
around the inner ($i$) and outer ($o$) electron pockets. When $\lambda_{\mathrm{SOC}}\ll\epsilon_{m}$,
as it is apparent from Fig.~\ref{Fig:IronFS}a, the outer electron
pocket consists mostly of $d_{xy}$ orbital spectral weight, whereas
the inner pocket consists mostly of $d_{xz}/d_{yz}$ spectral weigh.
The $d_{xz}$ and $d_{yz}$ gap functions have essentially the same
angular dependence as in the case without SOC, shown previously in
Fig.~\ref{Fig:NemSC}b. Consequently, the gap anisotropy depends
strongly on the ratio $\lambda_{1}/\lambda_{2}$ between the two nematic
couplings. For $\lambda_{1}\approx\lambda_{2}$, the gaps are nearly
isotropic around the inner and outer pockets, whereas for $\lambda_{1}<\lambda_{2}$
or $\lambda_{1}>\lambda_{2}$, the gaps are anisotropic in both pockets.

\begin{figure}[htbp]
\includegraphics[width=0.9\columnwidth]{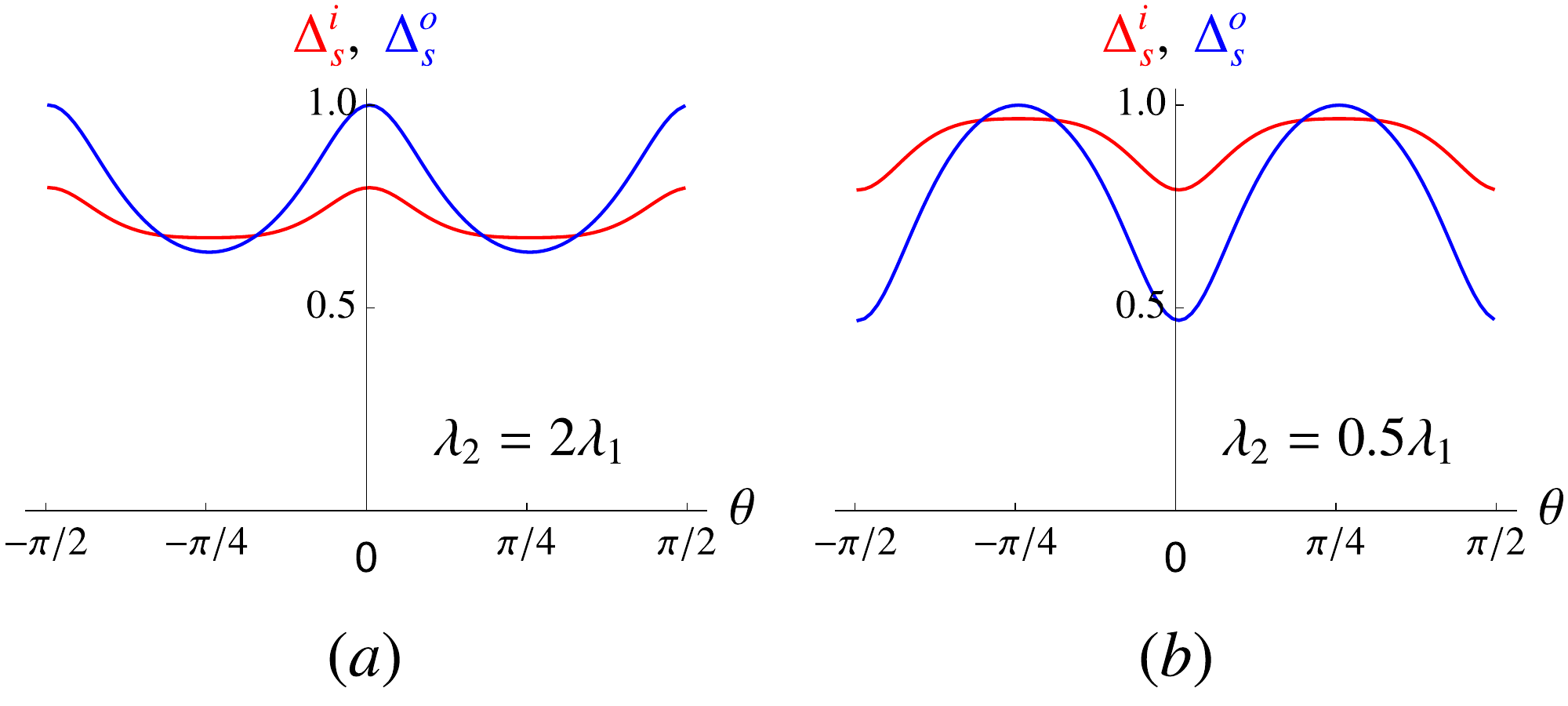} \protect\protect\caption{Angular dependence of the SC gap along the inner (red) and outer (blue)
electron pocket in the case where the SOC coupling is much larger
than the electron pockets mismatch. The positions of the gap minima
are controlled by $\lambda_{2}/\lambda_{1}$. \label{Fig:SCLargeMix} }
\end{figure}

The gap structure however changes dramatically in the case $\lambda_{\mathrm{SOC}}\gg\epsilon_{m}$
(with both still much smaller than the Fermi energy). In this case,
the two reconstructed electron pockets are fully hybridized, implying
that their orbital weights are similar. As a result, the SC gaps on
the inner and outer pockets are weakly anisotropic for all values
of the ratio $\lambda_{1}/\lambda_{2}$, whose main effect is to displace
the position of the gap maxima. While for $\lambda_{1}<\lambda_{2}$
the gap minima are located at the intersection points between the
two un-hybridized electron pockets, $\theta=\pm\pi/4$, for $\lambda_{1}>\lambda_{2}$
the gap minima are found at the intersection points (see Fig.~\ref{Fig:SCLargeMix}).
Interestingly, recent ARPES experiments in monolayer FeSe observe
gap maxima at $\theta=\pm\pi/4$~\cite{Zhang15}, whereas STM measurements
in the intercalated Li$_{1-x}$(OH)$_{x}$FeSe compound report gap
minima at $\theta=\pm\pi/4$~\cite{Wen16}.

Besides SOC, the inversion-symmetry breaking (ISB) at the interface
of thin films also lifts the degeneracy between $s$-wave and $d$-wave
in the case of FeSe thin films. In terms of the low-energy spinor
states, ISB gives rise to the term \cite{Hu14}:
\begin{equation}
H_{\mathrm{ISB}}=\lambda_{\mathrm{ISB}}\sum_{\fvec k}\Psi_{X}^{\dag}\frac{\tau_{0}+\tau_{3}}{2}\Psi_{Y}+h.c.\label{H_RSB}
\end{equation}

Similarly to SOC, $\lambda_{\mathrm{ISB}}$ hybridizes the two electron
pockets and favors $s$-wave over $d$-wave, lifting the degeneracy
between the two states (Fig.~\ref{fig_SC_SOC}b) and enhancing the
$s$-wave pairing instability. As shown in the SM, the effect of ISB
on the angular dependence of the gap functions around the inner and
outer pockets is very similar to the effect of SOC. The only difference
is that because ISB barely couples to the $d_{xy}$ orbitals, the
gaps remain moderately anisotropic.

So far we considered only the zero-momentum contribution of the nematic
fluctuations. In general, however, $\chi_{\mathrm{nem}}^{-1}\left(\mathbf{q}\right)=\xi_{\mathrm{nem}}^{-2}+q^{2}$.
Thus, although small-momentum fluctuations do not couple the $X$
and $Y$ pockets, leaving the $s$-wave/$d$-wave degeneracy intact,
large-momentum fluctuations couple them, giving rise to their own
free-energy coupling $\gamma'$ in Eq. (\ref{delta_F_SC}). As shown
in the SM, however, $\gamma'\ll\gamma$, implying that the small-momentum
approximation is sensible.

Besides SOC and ISB, other effects can lift the $s$-wave/$d$-wave
degeneracy promoted by the dominant nematic fluctuations. For instance,
magnetic fluctuations peaked at $(\pi,\pi)$ would favor the $d$-wave
state \cite{Khodas12,Hinojosa15}, whereas a momentum-independent
electron-phonon interaction would favor the $s$-wave state. To the
best of our knowledge, no sign of $(\pi,\pi)$ magnetic order has
been observed in FeSe thin films with only electron pockets. First-principle
calculations for the momentum-independent phonon coupling estimate
a resulting $T_{c}\lesssim1$ K \cite{Xing14}, an energy scale that
may be too small to significantly lift the degeneracy, since $T_{c}\approx40$
K in FeSe thin films.

Previous works have shown that forward-scattering phonons can lead
to a sizable enhancement of $T_{c}$ in FeSe films grown over SrTiO$_{3}$
or BaTiO$_{3}$~\cite{Lee12,Rademaker16,Johnston16,DHLee_STO,Dolgov16,Millis16}.
Indeed, the observation of replica band in ARPES measurements highlights
the importance of this phonon mode \cite{Shen14}. Similarly to the
nematic fluctuations studied here, forward-scattering phonons are
also peaked at zero momentum, and therefore are expected to also promote
degenerate $s$-wave/$d$-wave SC states \cite{Dolgov16}. In this
regard, the two pairing mechanisms may cooperate to promote a robust
SC state, whose degeneracy is lifted by SOC or ISB. While a detailed
analysis of this problem is beyond the scope of this work, it is tempting
to attribute to this cooperative effect the fact that $T_{c}$ is
higher in FeSe films grown over titanium oxide interfaces as compared
to other types of interfaces or other FeSe-based compounds.

\emph{Summary}\ \ \ In summary, we showed that the combined effect
of nematic fluctuations and SOC/ISB favors an $s$-wave state in electron-doped
thin films of FeSe, in agreement with recent experimental proposals
\cite{FengNP15}. The role played by SOC and ISB is fundamental to
lift the degeneracy with the sub-leading $d$-wave state, which suppresses
the onset of long-range SC order. Although nematic fluctuations are
momentum-independent in our model, the gap function can acquire a
pronounced angular dependence since the nematic order parameter couples
differently to $d_{xz}/d_{yz}$ and $d_{xy}$ orbitals. Interestingly,
in the regime where the SOC and ISB couplings are larger than the
mismatch between the electron pockets, we obtain a gap function whose
angular dependence agrees qualitatively with measurements in monolayer
FeSe and intercalated Li$_{1-x}$(OH)$_{x}$FeSe. More generally,
our work provides an interesting framework in which superconductivity
can develop in the presence of nematic fluctuations.
\begin{acknowledgments}
We thank A. Chubukov, S. Lederer, X. Liu, A. Millis, M. Khodas, S.
Kivelson, S. Raghu, D. Scalapino, M. Sch\"{u}t, Y. Wang, O. Vafek, and
Y. Y. Zhao for fruitful discussions. This work was supported by the
U.S. Department of Energy, Office of Science, Basic Energy Sciences,
under Award number DE-SC0012336. \end{acknowledgments}


\widetext \vspace{0.5cm}

\begin{center}
\textbf{\large{}{}{}Supplementary Material for ``Superconductivity
in FeSe thin films driven by the interplay between nematic fluctuations
and spin-orbit coupling{}''}{\large{}{} }
\par\end{center}

\setcounter{equation}{0} \setcounter{figure}{0} \setcounter{subfigure}{0}
\setcounter{table}{0} \makeatletter \global\long\def\theequation{S\arabic{equation}}
 \global\long\def\thefigure{S\arabic{figure}}
 \makeatother

\global\long\def\bibnumfmt#1{[S#1]}
 \global\long\def\citenumfont#1{S#1}

\section{Superconducting gap equations}

\subsection{No spin-orbit coupling}

The non-interacting Hamiltonian in terms of the spinors $\psi_{X,Y}$
is given by $\mathcal{H}_{0}=\sum\limits _{\fvec k,i=X,Y}\Psi_{i}^{\dagger}(\fvec k)\hat{H}_{i}(\fvec k)\Psi_{i}(\fvec k)$,
with:

\begin{align}
\hat{H}_{X} & =\begin{pmatrix}\epsilon_{1}+\frac{\fvec K^{2}}{2m_{1}}-a_{1}K_{x}K_{y} & -iv_{-}(\fvec K)\\
iv_{-}(\fvec K) & \epsilon_{3}+\frac{\fvec K^{2}}{2m_{3}}-a_{3}K_{x}K_{y}
\end{pmatrix}\ ,\quad\hat{H}_{Y}=\begin{pmatrix}\epsilon_{1}+\frac{\fvec K^{2}}{2m_{1}}+a_{1}K_{x}K_{y} & -iv_{+}(\fvec K)\\
iv_{+}(\fvec K) & \epsilon_{3}+\frac{\fvec K^{2}}{2m_{3}}+a_{3}K_{x}K_{y}
\end{pmatrix}\label{EqnS:Band}
\end{align}
and
\begin{equation}
v_{\pm}(\fvec K)=v(\pm K_{x}+K_{y})+p_{1}(\pm K_{x}^{3}+K_{y}^{3})+p_{2}K_{x}K_{y}(K_{x}\pm K_{y})\ .
\end{equation}
where $(K_{x},K_{y})$ refer to the 2-Fe Brillouin zone (BZ). The
parameters in the Hamiltonian $\mathcal{H}_{0}$ are taken from Table
IX of Ref.~\cite{Oskar}. The SC gap equations are given by the Feynman
diagram in Fig.~\ref{FigS:NemSCSum}:
\begin{equation}
\eta_{\alpha}\hat{M}_{\alpha}=\chi_{\mathrm{nem}}T\sum_{n,\fvec k}\left(\hat{\lambda}^{\mathrm{nem}}\right)^{T}\hat{G}_{-k}^{T}\hat{M}_{\alpha}\hat{G}_{k}\hat{\lambda}^{\mathrm{nem}}\ .\label{EqnS:GapEqn}
\end{equation}
where $\hat{\lambda}$, $\hat{G}_{K}$ are $4\times4$ matrices:
\[
\hat{\lambda}^{\mathrm{nem}}=\begin{pmatrix}\hat{\lambda}_{X}^{\mathrm{nem}} & 0\\
0 & \hat{\lambda}_{Y}^{\mathrm{nem}}
\end{pmatrix}\ ,\quad\hat{G}_{K}=\big(i\omega_{n}I_{4\times4}-H(\fvec K)\big)^{-1}\ \text{with }\ \hat{H}(\fvec K)=\begin{pmatrix}\hat{H}_{X}(\fvec K) & 0\\
0 & \hat{H}_{Y}(\fvec K)
\end{pmatrix}
\]
and $\hat{\lambda}_{X}^{\mathrm{nem}}$ and $\hat{\lambda}_{X}^{\mathrm{nem}}$
are defined as in the main text. For simplicity, we introduce two
parameters to describe the nematic couplings $\lambda_{1}$ and $\lambda_{2}$:
$\lambda=\sqrt{\lambda_{1}^{2}+\lambda_{2}^{2}}$ and $\chi=\tan^{-1}(\lambda_{2}/\lambda_{1})$.
$\hat{M}_{\alpha}$ is also a $4\times4$ matrix with the label $\alpha$
referring to different irreducible representations of the $P4/nmm$
space group. Following Ref.~\cite{Oskar}, there are 6 different
irreducible representations corresponding to singlet pairing at the
electron pockets (namely, $A_{1g}$, $B_{2g}$, $A_{2u}$, $B_{2u}$,
$E_{g}$, and $E_{u}$):
\begin{align}
 & \hat{M}_{A_{1g}/B_{2g}}=\begin{pmatrix}\Delta_{1}\\
 & \Delta_{2}\\
 &  & \pm\Delta_{1}\\
 &  &  & \pm\Delta_{2}
\end{pmatrix}\quad\hat{M}_{A_{2u}/B_{2u}}=\begin{pmatrix}0 & \dfrac{\tau_{0}\pm\tau_{3}}{2}\\
\dfrac{\tau_{0}\pm\tau_{3}}{2} & 0
\end{pmatrix}\quad\hat{M}_{E_{g}^{(1)}}=\begin{pmatrix}0 & \tau_{-}\\
\tau_{+} & 0
\end{pmatrix}\nonumber \\
 & \hat{M}_{E_{g}^{(2)}}=\begin{pmatrix}0 & -\tau_{+}\\
-\tau_{-} & 0
\end{pmatrix}\quad\hat{M}_{E_{u}^{(1)}}=\begin{pmatrix}0 & 0\\
0 & \tau_{1}
\end{pmatrix}\quad\hat{M}_{E_{u}^{(2)}}=\begin{pmatrix}\tau_{1} & 0\\
0 & 0
\end{pmatrix}
\end{align}
Note that because $E_{g}$ and $E_{u}$ are two-dimensional representations,
we introduced the superscripts $(1)$ and $(2)$ for the two components
of the representation that give the same eigenvalue $\eta$. To compute
$T_{c}$, we note that an infinitesimal pairing field $\Delta_{0}$
is renormalized by nematic fluctuations according to the diagrammatic
series shown in Fig.~\ref{FigS:NemSCSum}:

\begin{figure}[htbp]
\includegraphics[scale=0.9]{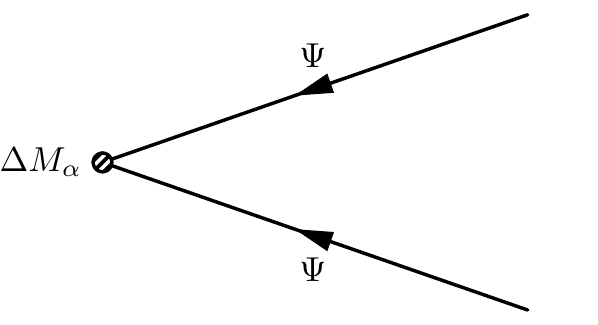}  \raisebox{1.3cm}{\noindent\LARGE$=$}
\includegraphics[scale=0.9]{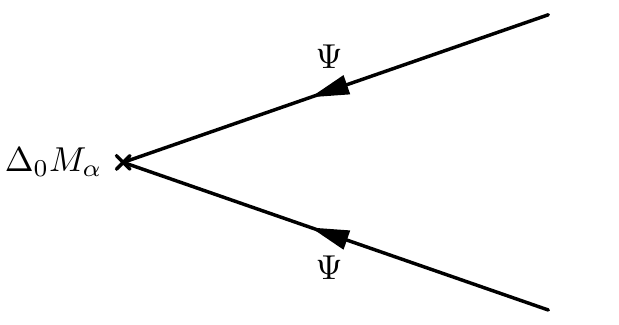} \raisebox{1.3cm}{\noindent\LARGE$+$} \includegraphics[scale=0.9]{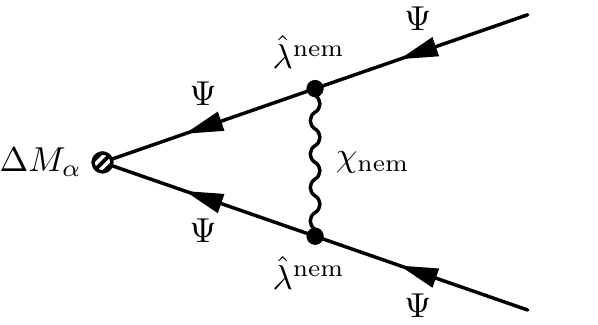}
\caption{The effective pairing field $\Delta M_{\alpha}$ is the sum of the
infinitesimal pairing field $\Delta_{0}M_{\alpha}$ and the effective
pairing field dressed by nematic fluctuations.}
\label{FigS:NemSCSum}
\end{figure}

\begin{equation}
\Delta\hat{M}_{\alpha}=\Delta_{0}\hat{M}_{\alpha}+\Delta\chi_{\mathrm{nem}}T\sum_{n,\fvec k}\left(\hat{\lambda}^{\mathrm{nem}}\right)^{T}\hat{G}_{-k}^{T}\hat{M}_{\alpha}\hat{G}_{k}\hat{\lambda}^{\mathrm{nem}}\quad\Longrightarrow\quad\Delta=\frac{\Delta_{0}}{1-\eta_{\alpha}}
\end{equation}

\begin{figure}[htbp]
\centering \subfigure[\label{Fig:SCPair:Eigen}]{\includegraphics[scale=0.7]{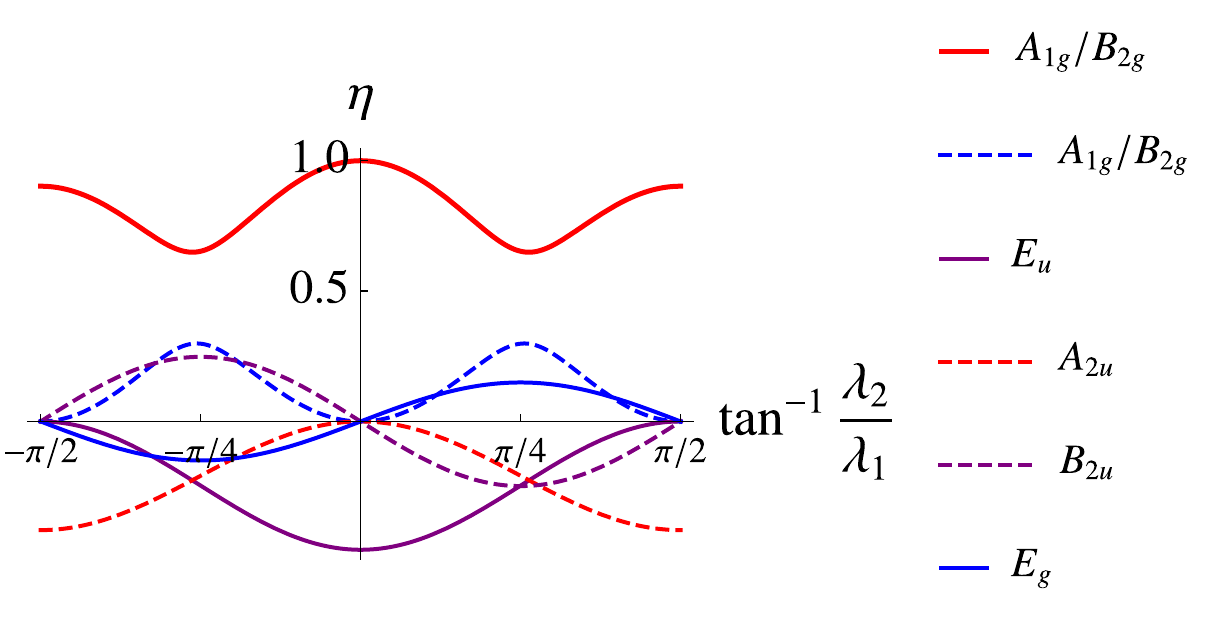}}\hfill{}
\subfigure[\label{Fig:SCPair:Sol}]{\includegraphics[scale=0.6]{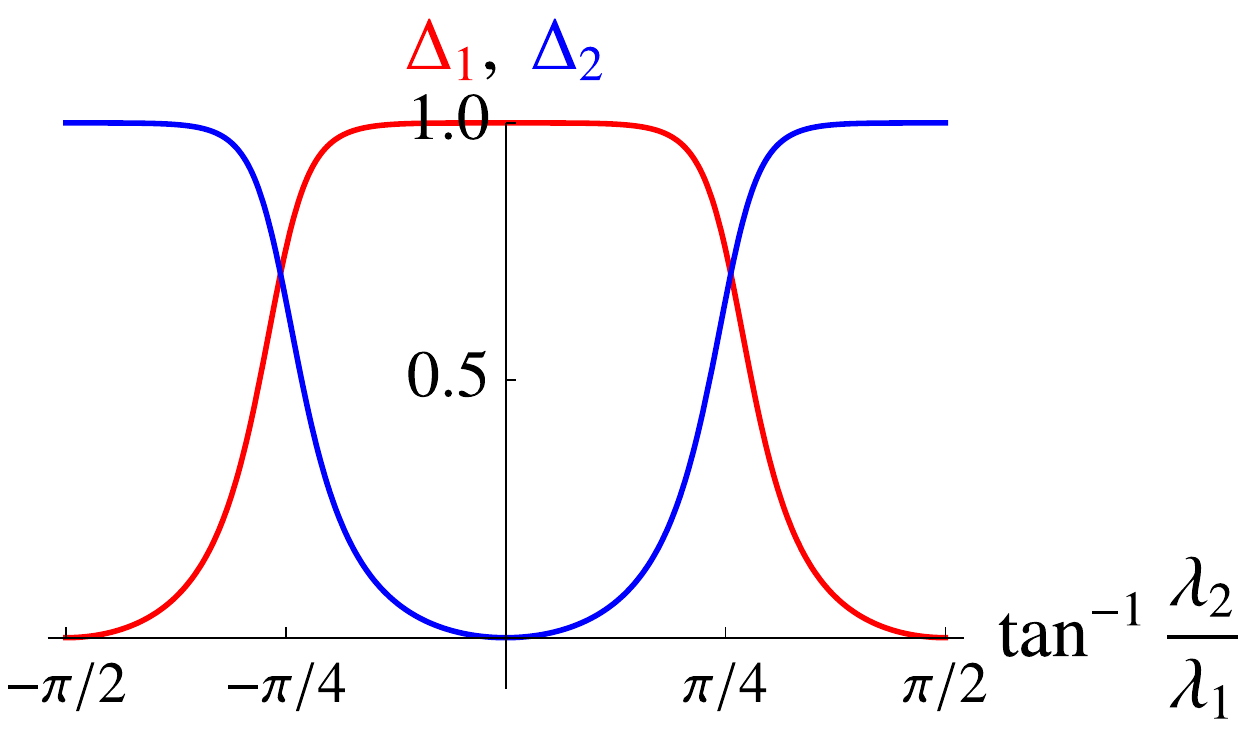}}
\protect\protect\protect\protect\protect\caption{(a) The eigenvalues for different SC channels without SOC or ISB,
as function of the ratio $\lambda_{2}/\lambda_{1}$ between the two
nematic coupling constants. (b) The two intra-orbital gap functions
$\Delta_{1}$ and $\Delta_{2}$ corresponding to the solution of the
gap equations in the degenerate $A_{1g}$ /$B_{2g}$ channel.}

\label{Fig:SCPair}
\end{figure}

Therefore,~$T_{c}$ is obtained when the largest eigenvalue $\eta_{\alpha}(T=T_{c})=1$.
In our paper, the value of the coupling $\lambda^{2}\chi_{\mathrm{nem}}$
is set such that $T_{c}=5$ meV for the case of $A_{1g}/B_{2g}$ pairing
when $\lambda_{2}=0$, i.e. $\chi=0$. Fig.~\ref{Fig:SCPair} shows
the other eigenvalues $\eta_{\alpha}$ at the same temperature $T=5$
meV. Clearly, the $A_{1g}$ and $B_{1g}$ states are the degenerate
leading SC instabilities of this system. The corresponding values
for $\Delta_{1}$ and $\Delta_{2}$ as a function of $\chi=\tan^{-1}(\lambda_{2}/\lambda_{1})$
is also shown in the figure.

To project the gaps onto the Fermi pockets, $\Delta_{X/Y}(\fvec K)$,
we introduce the spinor $u_{X/Y}\left(\mathbf{K}\right)=\big(u_{X/Y,1}(\fvec K),u_{X/Y,2}(\fvec K)\big)^{T}$
that diagonalizes $\hat{H}_{X/Y}$ and whose eigenvalue corresponds
to the band dispersion that crosses the Fermi level. Then, the projected
gap is given by:
\begin{equation}
\Delta_{X}(\fvec K)=\Delta_{1}\big|u_{X,1}(\fvec K)\big|^{2}+\Delta_{2}\big|u_{X,2}(\fvec K)\big|^{2}
\end{equation}

\subsection{Non-zero spin-orbit coupling}

The SOC interaction is given by:
\begin{equation}
\mathcal{H}_{\mathrm{SOC}}=\frac{i}{2}\lambda_{\mathrm{SOC}}\Psi_{Y}^{\dag}\left(\tau_{+}\otimes\sigma_{1}+\tau_{-}\otimes\sigma_{2}\right)\Psi_{X}+h.c.
\end{equation}

In the presence of SOC, spin-singlet and spin-triplet pairings are
mixed. In general, the latter can be written as $\Psi^{T}\fvec M\otimes i\sigma_{2}\fvec\sigma\Psi$,
where $\Psi$ is the eight-component spinor $\Psi=(\Psi_{X\sigma}\ ,\ \Psi_{Y\sigma})^{T}$.
Since the spin component is symmetric for spin-triplet pairing, the
orbital part $\fvec M$ must be anti-symmetric. Since $A_{1g}$ and
$B_{2g}$ are the two leading SC instabilities in the absence of SOC,
we only focus on these two channels here. These irreducible representations
can only be obtained if both the spin component and the orbital component
transform as $E_{g}$, since $E_{g}\otimes E_{g}=A_{1g}\oplus A_{2g}\oplus B_{1g}\oplus B_{2g}$.
The spin combination that transforms according to $E_{g}$ is $(i\sigma_{2}\sigma_{1},i\sigma_{2}\sigma_{2})$;
for the orbital part, which must be anti-symmetric (i.e. $E_{g}^{-}$),
we have:
\begin{equation}
-i\big(\Psi_{Y}^{T}\tau_{+}\Psi_{X}-\Psi_{X}^{T}\tau_{-}\Psi_{Y},\Psi_{Y}^{T}\tau_{-}\Psi_{X}-\Psi_{X}^{T}\tau_{+}\Psi_{Y}\big)
\end{equation}
which corresponds to inter-pocket pairing. Writing it in the form
$\big(\Psi^{T}M_{1}\Psi,\Psi^{T}M_{2}\Psi\big)$, we readily obtain
the non-zero matrix elements $\big(M_{2}\big)_{14}=-\big(M_{2}\big)_{41}=\big(M_{1}\big)_{23}=-\big(M_{1}\big)_{32}=i$.
Combined with the spin part, we obtain the following gap functions:
\begin{align}
A_{1g}: & \qquad\Psi^{T}\left(\begin{pmatrix}\Delta_{1}^{A} & 0 & 0 & 0\\
0 & \Delta_{2}^{A} & 0 & 0\\
0 & 0 & \Delta_{1}^{A} & 0\\
0 & 0 & 0 & \Delta_{2}^{A}
\end{pmatrix}\otimes i\sigma_{2}+\begin{pmatrix}0 & 0 & 0 & 0\\
0 & 0 & i\Delta_{3}^{A} & 0\\
0 & -i\Delta_{3}^{A} & 0 & 0\\
0 & 0 & 0 & 0
\end{pmatrix}\otimes\sigma_{3}+\begin{pmatrix}0 & 0 & 0 & -\Delta_{3}^{A}\\
0 & 0 & 0 & 0\\
0 & 0 & 0 & 0\\
\Delta_{3}^{A} & 0 & 0 & 0
\end{pmatrix}\otimes\sigma_{0}\right)\Psi\\
B_{2g}: & \qquad\Psi^{T}\left(\begin{pmatrix}\Delta_{1}^{B} & 0 & 0 & 0\\
0 & \Delta_{2}^{B} & 0 & 0\\
0 & 0 & -\Delta_{1}^{B} & 0\\
0 & 0 & 0 & -\Delta_{2}^{B}
\end{pmatrix}\otimes i\sigma_{2}+\begin{pmatrix}0 & 0 & 0 & 0\\
0 & 0 & i\Delta_{3}^{B} & 0\\
0 & -i\Delta_{3}^{B} & 0 & 0\\
0 & 0 & 0 & 0
\end{pmatrix}\otimes\sigma_{3}+\begin{pmatrix}0 & 0 & 0 & \Delta_{3}^{B}\\
0 & 0 & 0 & 0\\
0 & 0 & 0 & 0\\
-\Delta_{3}^{B} & 0 & 0 & 0
\end{pmatrix}\otimes\sigma_{0}\right)\Psi
\end{align}

\begin{figure}[htbp]
\centering \subfigure[\label{Fig:SCSOCPair:A1gSol}]{\includegraphics[scale=0.5]{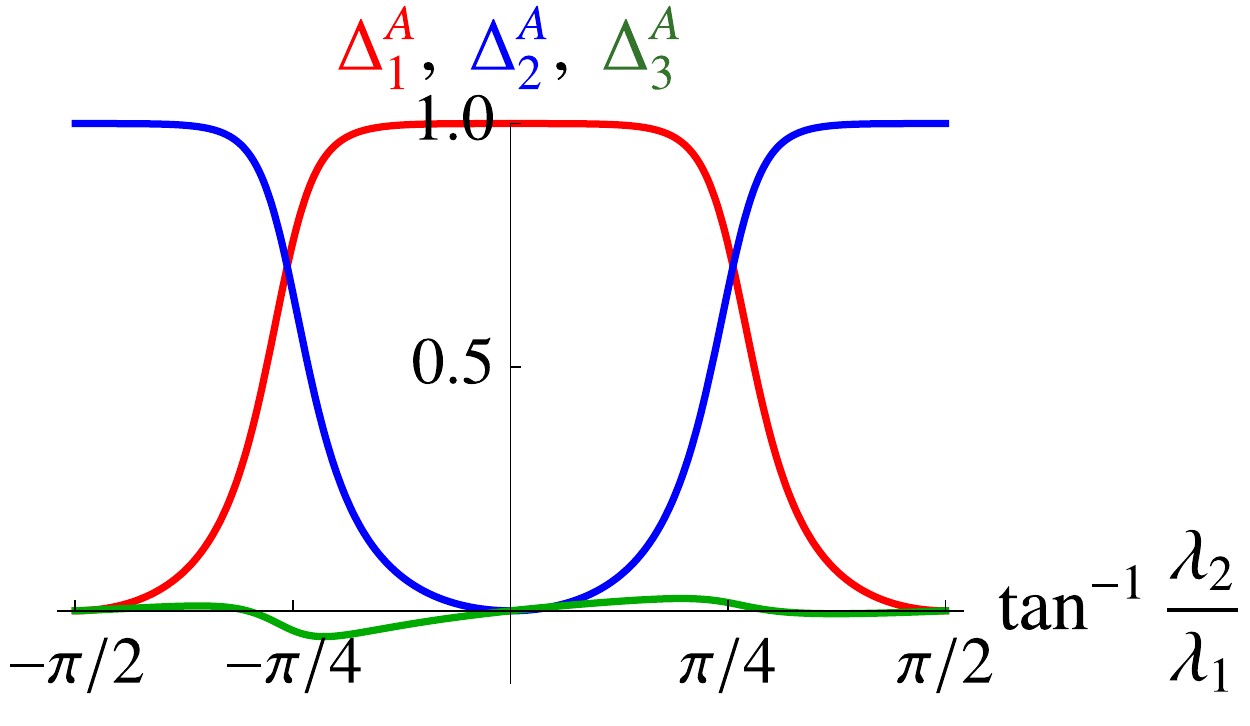}}\hspace{1cm}
\subfigure[\label{Fig:SCSOCPair:B2gSol}]{\includegraphics[scale=0.5]{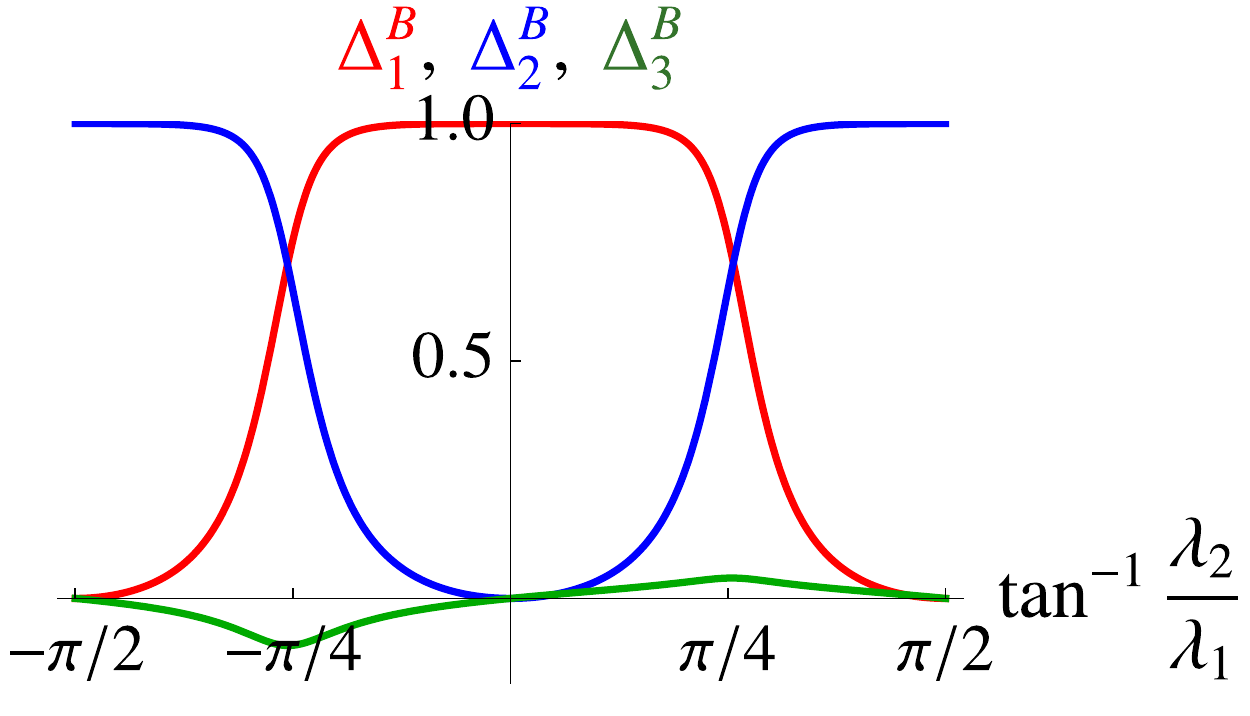}}
\protect\protect\protect\protect\protect\caption{The solution of the $A_{1g}$ and $B_{2g}$ pairing gaps in the presence
of SOC. The two nematic couplings are given by $\lambda_{1}$ and
$\lambda_{2}$.}

\label{Fig:SCSOCPair}
\end{figure}

The three gaps $\Delta_{1}$, $\Delta_{2}$, and $\Delta_{3}$ are
determined by solving the gap equation. In Fig. \ref{Fig:SCSOCPair},
we show the solution of the gap equations: clearly, the presence of
spin-orbit coupling lifts the degeneracy between $A_{1g}$ and $B_{2g}$,
as illustrated in Fig.~2 of the main text. Here we used $\lambda_{\mathrm{SOC}}=50$meV.
We also note that the admixture with the triplet component is small.

To calculate the momentum dependence of the gap function, we project
the gaps $\Delta_{1}$, $\Delta_{2}$, and $\Delta_{3}$ along the
Fermi surface. Since the Hamiltonian does not break time reversal
symmetry and inversion symmetry, each band is doubly degenerate (Kramers
degeneracy). To show this more clearly, we define two new 4-component
spinors related by time reversal symmetry:
\begin{equation}
\Phi_{1}=\big(\Psi_{X\uparrow}\quad\Psi_{Y\downarrow}\big)\ ,\quad\mbox{and}\quad\Phi_{2}=\big(\Psi_{X\downarrow}\quad-\Psi_{Y\uparrow}\big)
\end{equation}

Both $\mathcal{H}_{0}$ and the SOC term are diagonal in this representation:
\[
\mathcal{H}_{0}+\mathcal{H}_{\mathrm{SOC}}=\sum_{\fvec K}\Phi_{1}^{\dag}\hat{H}_{1}\Phi_{1}+\sum_{\fvec K}\Phi_{2}^{\dag}\hat{H}_{2}\Phi_{2}\ ,\quad\mbox{with }\quad\hat{H}_{1}=\begin{pmatrix}\hat{H}_{X} & \hat{h}\\
\hat{h}^{\dag} & \hat{H}_{Y}
\end{pmatrix}\ ,\mbox{\,\ }H_{2}=\begin{pmatrix}\hat{H}_{X} & \hat{h}^{*}\\
\hat{h}^{T} & \hat{H}_{Y}
\end{pmatrix}\quad,\mbox{and }\hat{h}=\frac{\lambda_{\mathrm{SOC}}}{2}\begin{pmatrix}0 & -1\\
-i & 0
\end{pmatrix}
\]

It follows immediately that $H_{1}(\fvec K)=H_{2}^{*}(-\fvec K)$.
Furthermore, because the system is also invariant under inversion,
the energy dispersions of $\Phi_{1}$ and $\Phi_{2}$ are exactly
the same. This implies that each band in the system is doubly degenerate,
although neither $\Phi_{1}$ nor $\Phi_{2}$ has degeneracies. Upon
diagonalization, we find that the two overlapping electron pockets
split due to the SOC, forming an inner and an outer electron pocket.

\begin{figure}[htbp]
\centering \subfigure[\label{Fig:SCSOCBand:InA1g}]{\includegraphics[scale=0.5]{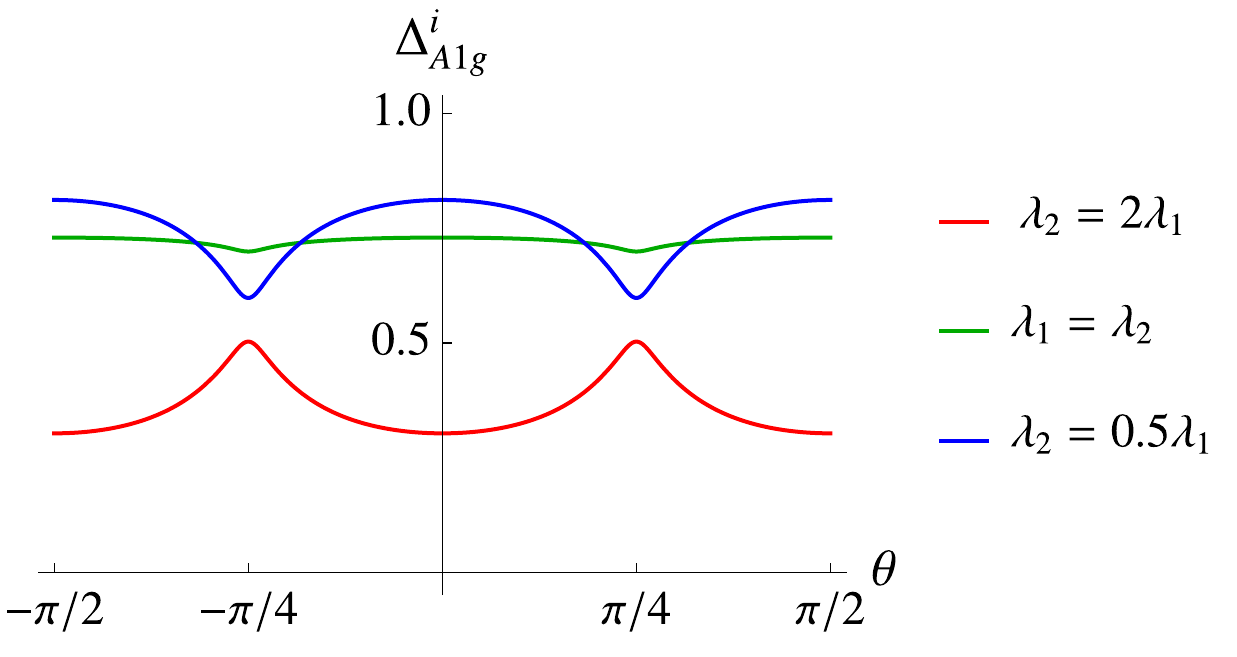}}
\subfigure[\label{Fig:SCSOCBand:OutA1g}]{\includegraphics[scale=0.5]{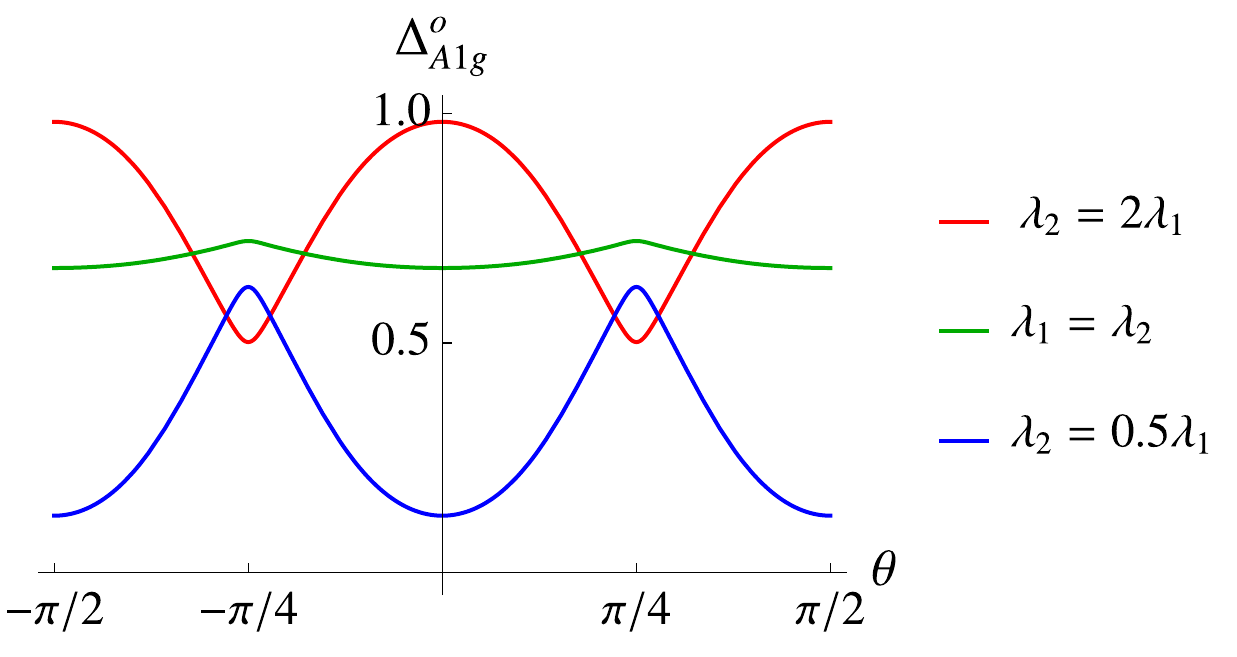}}
\subfigure[\label{Fig:SCSOCBand:InB2g}]{\includegraphics[scale=0.5]{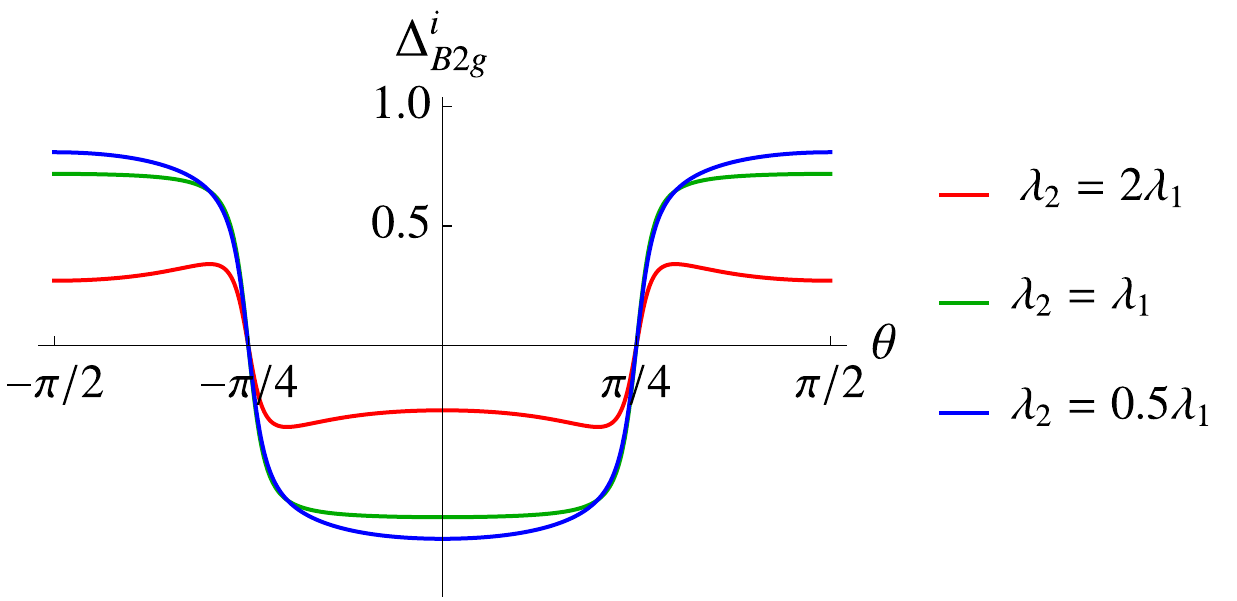}}
\subfigure[\label{Fig:SCSOCBand:OutB2g}]{\includegraphics[scale=0.5]{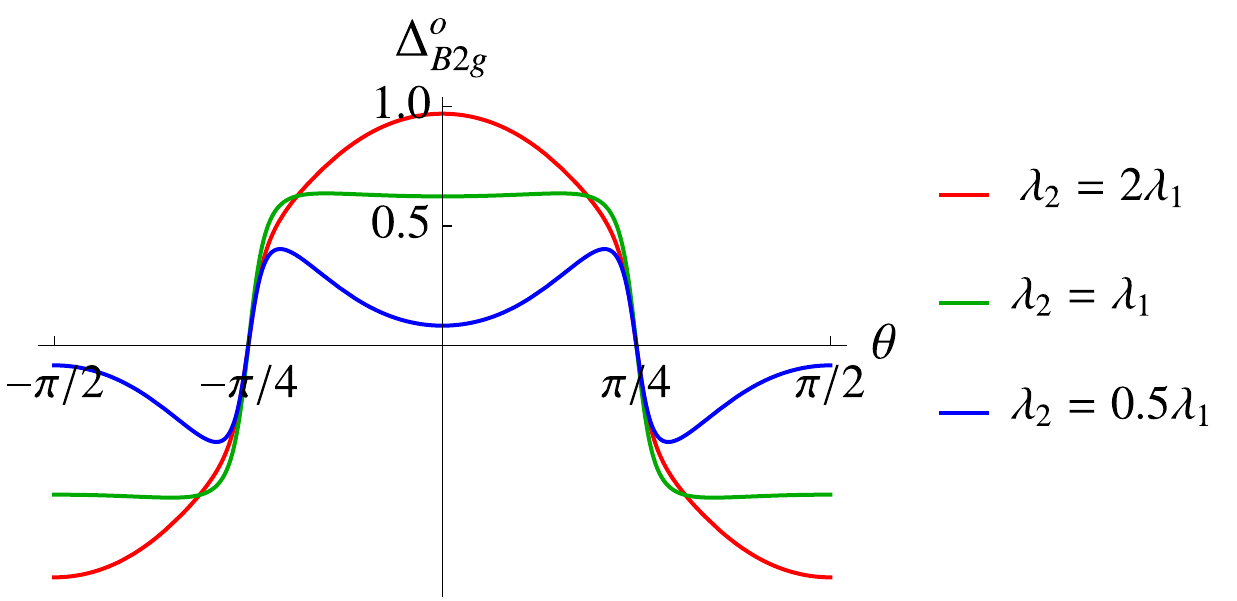}}
\protect\protect\protect\protect\protect\caption{Projected gap on the inner ($i$) and outer ($o$) electron pockets,
with $\lambda_{\mathrm{SOC}}=50$meV. (a) and (b) correspond to the
$A_{1g}$ pairing channel, whereas (c) and (d) correspond to the $B_{2g}$
channel. The two nematic couplings are given by $\lambda_{1}$ and
$\lambda_{2}$.}

\label{Fig:SCSOCBand}
\end{figure}

As for the pairing interaction, we find that both $A_{1g}$ and $B_{2g}$
gaps couple the two spinors, $\Phi_{1}^{T}\hat{\Delta}_{A_{1g}}\Phi_{2}$
and $\Phi_{1}^{T}\hat{\Delta}_{B_{2g}}\Phi_{2}$, with
\[
\hat{\Delta}_{A_{1g}}=\begin{pmatrix}\Delta_{1}^{A} & 0 & 0 & \Delta_{3}^{A}\\
0 & \Delta_{2}^{A} & -i\Delta_{3}^{A} & 0\\
0 & i\Delta_{3}^{A} & \Delta_{1}^{A} & 0\\
\Delta_{3}^{A} & 0 & 0 & \Delta_{2}^{A}
\end{pmatrix}\ ,\mbox{and }\quad\hat{\Delta}_{B_{2g}}=\begin{pmatrix}\Delta_{1}^{B} & 0 & 0 & -\Delta_{3}^{B}\\
0 & \Delta_{2}^{B} & -i\Delta_{3}^{B} & 0\\
0 & i\Delta_{3}^{B} & -\Delta_{1}^{B} & 0\\
-\Delta_{3}^{B} & 0 & 0 & -\Delta_{2}^{B}
\end{pmatrix}
\]

This allows us to project the gap onto the Fermi surface in a straightforward
way. Consider for instance the $A_{1g}$ gap projected onto the inner
Fermi pocket. We first diagonalize $H_{1}(\fvec K)$ and $H_{2}\left(\fvec K\right)$
to obtain the band operators $u_{i,\fvec K}$ and $v_{i,\fvec K}$,
respectively. They are related to the orbital operators $d_{\mu,\fvec K}$
according to:
\begin{align}
d_{\mu,\fvec K} & =\sum_{i}\langle\mu|u_{i}(\fvec K)\rangle u_{i,\fvec K}\\
d_{\mu,\fvec K} & =\sum_{i}\langle\mu|v_{i}(\fvec K)\rangle v_{i,\fvec K}
\end{align}

Due to time-reversal symmetry, $\langle\mu|v_{i}(-\fvec K)\rangle=\langle u_{i}(\fvec K)|d_{\mu}\rangle$.
The SC Hamiltonian then becomes:
\begin{align}
\sum_{\mu\nu}\Delta_{\mu\nu}d_{\mu,-\fvec K}d_{\nu,\fvec K} & =\sum_{\mu\nu}\Delta_{\mu\nu}\sum_{ij}\langle\mu|v_{i}(-\fvec K)\rangle\langle\nu|u_{j}(\fvec K)\rangle v_{i,-\fvec K}u_{j,\fvec K}\nonumber \\
 & =\sum_{\mu\nu}\Delta_{\mu\nu}\sum_{ij}\langle u_{i}(\fvec K)|\mu\rangle\langle\nu|u_{j}(\fvec K)\rangle v_{i,-\fvec K}u_{j,\fvec K}
\end{align}

Projecting onto band $l$, we find:
\begin{equation}
\sum_{\mu\nu}\Delta_{\mu\nu}\langle u_{l}(\fvec K)|\mu\rangle\langle\nu|u_{l}(\fvec K)\rangle v_{l}(-\fvec K)u_{l}(\fvec K)=\langle u_{l}(\fvec K)|\hat{\Delta}|u_{l}(\fvec K)\rangle v_{l}(-\fvec K)u_{l}(\fvec K)
\end{equation}

As a result, the gap along the pocket corresponding to band $l$ given
by:
\begin{equation}
\Delta_{l}\left(\fvec K\right)=\langle u_{l}(\fvec K)|\hat{\Delta}|u_{l}(\fvec K)\rangle\ ,
\end{equation}
yielding the results shown in Fig.~\ref{Fig:SCSOCBand}.

\subsection{Inversion symmetry-breaking}

\begin{figure}[htbp]
\centering \subfigure[\label{Fig:InvSC:Eigen}]{\includegraphics[scale=0.6]{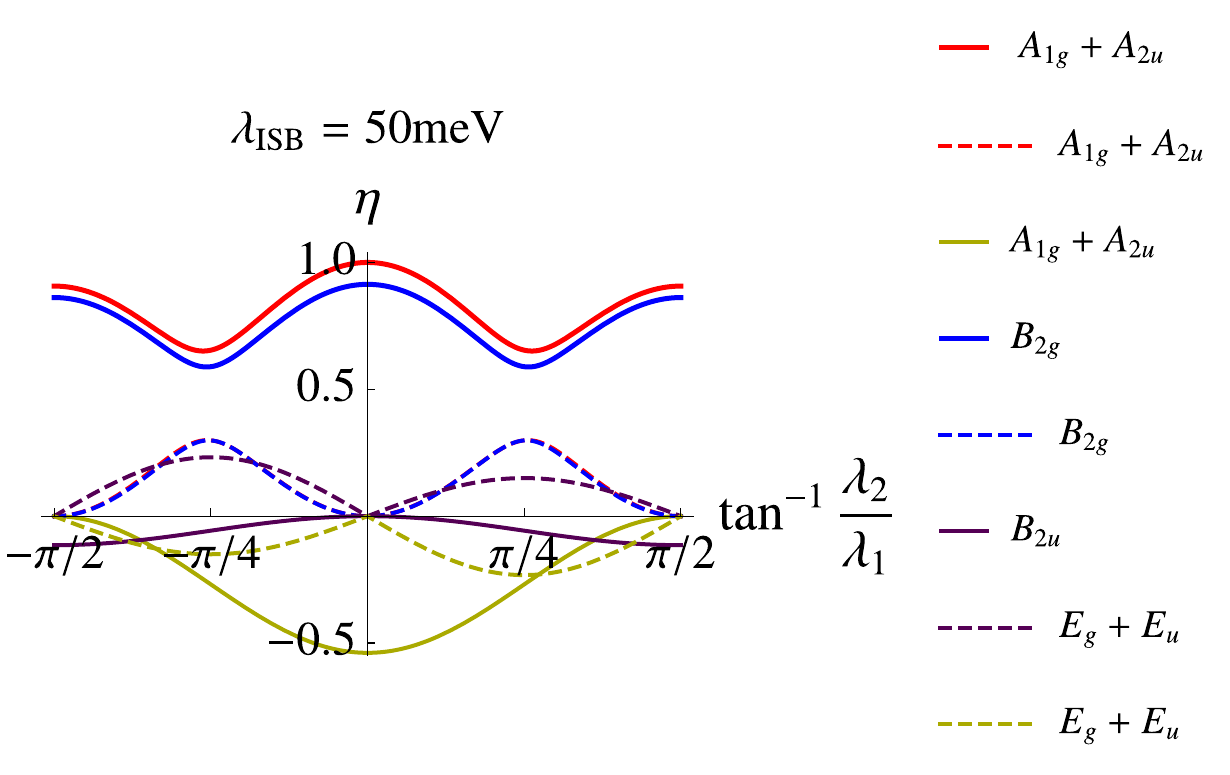}}
\subfigure[\label{Fig:InvSC:InA1g}]{\includegraphics[scale=0.6]{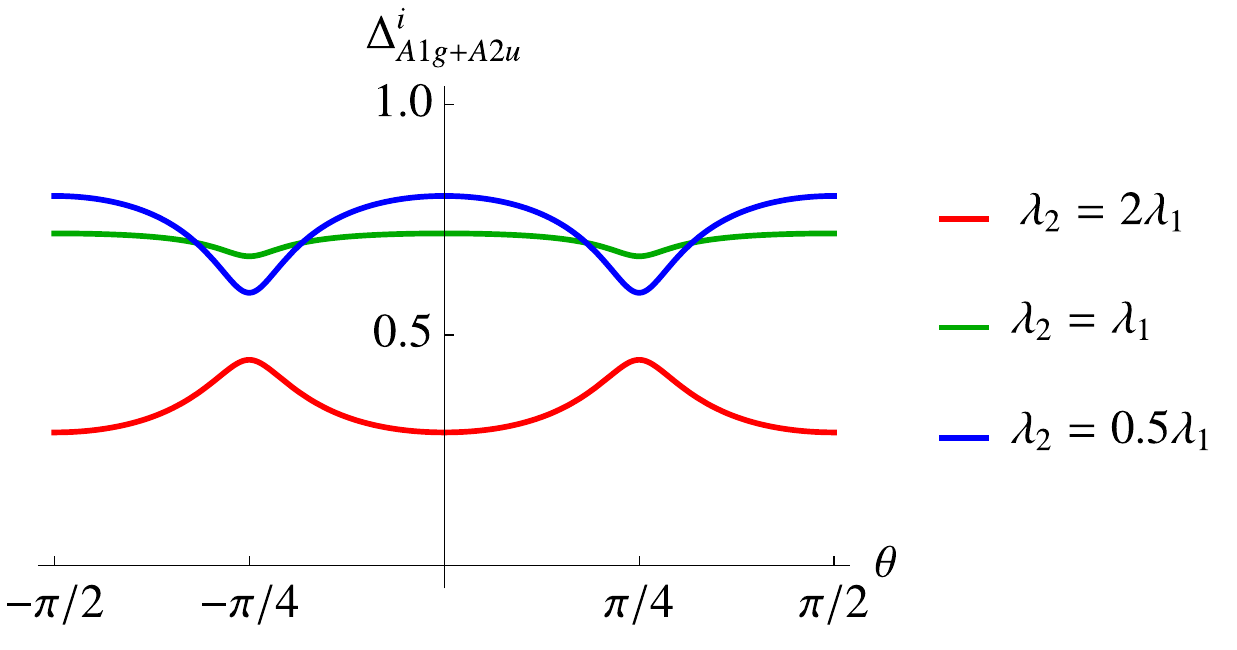}}
\subfigure[\label{Fig:InvSC:OutA1g}]{\includegraphics[scale=0.6]{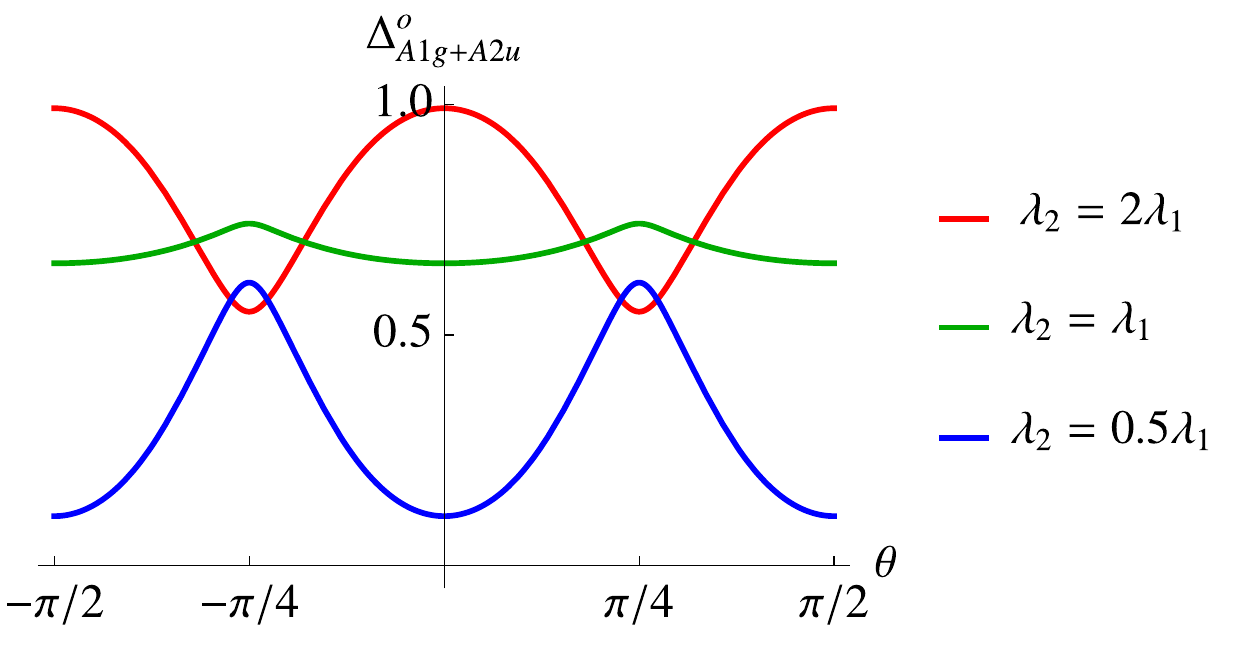}}
\protect\protect\protect\protect\protect\caption{(a) The SC eigenvalue $\eta$ in different SC channels for $\lambda_{\mathrm{ISB}}=50$meV.
(b) and (c): Projected gap on the inner ($i$) and outer ($o$) electron
pockets in the leading $A_{1g}+A_{2u}$ pairing channel. The two nematic
couplings are given by $\lambda_{1}$ and $\lambda_{2}$.}

\label{Fig:InvSC}
\end{figure}

As discussed in the main text, the inversion symmetry is broken at
the interface of thin films of FeSe. Considering the generators of
the $P4/nmm$ group, $\{\sigma_{z}|\half\half\}$ is the only symmetry
transformation broken, as it corresponds to a reflection $\sigma_{z}$
with respect to the Fe plane followed by a translation by $\left(\frac{1}{2},\frac{1}{2}\right)$
in the 2-Fe unit cell. Because the term:
\begin{equation}
H_{\mathrm{ISB}}=\lambda_{\mathrm{ISB}}\Psi_{X}^{\dagger}\frac{\tau_{0}+\tau_{3}}{2}\Psi_{Y}+h.c.
\end{equation}
acquires a minus sign upon the symmetry transformation $\{\sigma_{z}|\half\half\}$,
it must be generated once inversion symmetry is broken. Similarly
to the SOC term, the ISB term hybridizes the $X$ and $Y$ pockets,
and splits the degeneracy between the $A_{1g}$ and $B_{2g}$ pairing
states. Unlike the SOC case, there is no admixture with triplet components.
However, to solve the gap equations, one needs to introduce an admixture
with the pairing channel $A_{2u}$, resulting in the pairing matrix
of the form:
\begin{equation}
\hat{M}_{A_{1g}+A_{2u}}=\begin{pmatrix}\Delta_{1} & 0 & \Delta_{3} & 0\\
0 & \Delta_{2} & 0 & 0\\
\Delta_{3} & 0 & \Delta_{1} & 0\\
0 & 0 & 0 & \Delta_{2}
\end{pmatrix}
\end{equation}
In Fig. \ref{Fig:InvSC}, we show the solution of the gap equations
and the corresponding gap functions projected onto the Fermi surface.

\section{Superconducting free energy}

\subsection{No spin-orbit coupling}

In the absence of SOC and ISB, the $s$-wave and $d$-wave channels
have the same $T_{c}$. In terms of a free energy expansion, this
implies that, to quadratic order in the gaps:
\begin{equation}
F^{(2)}=\alpha\left(|\Delta_{s}|^{2}+|\Delta_{d}|^{2}\right)
\end{equation}

In this section we study how this degeneracy is affected by quartic
coefficients in the free energy, which go beyond the linearized SC
gap equations. For notation convenience, we define the superconducting
order parameters $\Delta_{s}$ and $\Delta_{d}$ by
\[
\hat{\Delta}_{A_{1g}}=\Delta_{s}\begin{pmatrix}\cos\alpha_{s} & 0 & 0 & 0\\
0 & \sin\alpha_{s} & 0 & 0\\
0 & 0 & \cos\alpha_{s} & 0\\
0 & 0 & 0 & \sin\alpha_{s}
\end{pmatrix}\otimes i\sigma_{2}\qquad\hat{\Delta}_{B_{2g}}=\Delta_{d}\begin{pmatrix}\cos\alpha_{d} & 0 & 0 & 0\\
0 & \sin\alpha_{d} & 0 & 0\\
0 & 0 & -\cos\alpha_{d} & 0\\
0 & 0 & 0 & -\sin\alpha_{d}
\end{pmatrix}\otimes i\sigma_{2}\ .
\]

They are related to the gaps $\Delta_{1,2}^{A/B}$ according to $\Delta_{1}^{A/B}=\Delta_{s/d}\cos\alpha_{s/d}$
and $\Delta_{2}^{A/B}=\Delta_{s/d}\sin\alpha_{s/d}$. The parameters
$\alpha_{s/d}$ are obtained from the gap equations. The quartic terms
of the free energy are given by:
\[
F^{(4)}=\frac{T}{2}\sum_{k}\mathrm{Tr}\Big(\hat{G}_{k}\hat{\Delta}\hat{G}_{-k}^{T}\hat{\Delta}^{\dagger}\hat{G}_{k}\hat{\Delta}\hat{G}_{-k}^{T}\hat{\Delta}^{\dagger}\Big)\ ,\quad\text{with }\quad\hat{\Delta}=\hat{\Delta}_{A_{1g}}+\hat{\Delta}_{B_{2g}}\ .
\]
yielding:
\begin{equation}
F^{(4)}=\beta_{1}|\Delta_{d}|^{4}+\gamma_{1}|\Delta_{s}|^{2}|\Delta_{d}|^{2}+\frac{\gamma_{2}}{2}\left(\Delta_{s}^{*2}\Delta_{d}^{2}+c.c.\right)+\beta_{2}|\Delta_{d}|^{4}\ ,
\end{equation}

Evaluating the trace gives $\gamma_{1}=2\gamma_{2}=4\beta_{1}=4\beta_{2}=4\beta$.
Thus, the free energy can be written as
\begin{equation}
F=\alpha\left(|\Delta_{s}|^{2}+|\Delta_{d}|^{2}\right)+\beta\left(|\Delta_{s}|^{4}+|\Delta_{d}|^{4}+4|\Delta_{s}|^{2}|\Delta_{d}|^{2}+\Delta_{s}\Delta_{d}^{*}+\Delta_{s}^{*}\Delta_{d}\right)
\end{equation}

Minimizing the free energy with respect to the relative phase between
$\Delta_{s}$ and $\Delta_{d}$ give $\pi/2$. Under this condition,
the free energy is given by:
\begin{equation}
F=\alpha\left(|\Delta_{s}|^{2}+|\Delta_{d}|^{2}\right)+\beta\left(|\Delta_{s}|^{2}+|\Delta_{d}|^{2}\right)^{2}
\end{equation}

Thus, besides the $U(1)$ symmetry related to the global phase, there
is an additional $U(1)$ symmetry related to the fact that only the
value of $|\Delta_{s}|^{2}+|\Delta_{d}|^{2}$ is fixed by minimization
of the free energy. Equivalently, we can write the $s$ and $d$ gaps
in terms of the gaps on the two electron pockets, $\Delta_{X}$ and
$\Delta_{Y}$. To see this, we note that the solution of the gap equations
for the $s$ and $d$ gaps give the same parameter $\alpha_{s}=\alpha_{d}$.
Then the total gap can be written as:

\begin{equation}
\hat{\Delta}=\hat{\Delta}_{A_{1g}}+\hat{\Delta}_{B_{2g}}=\left(\begin{array}{cc}
\left(\Delta_{s}+\Delta_{d}\right)\tilde{\tau} & 0\\
0 & \left(\Delta_{s}-\Delta_{d}\right)\tilde{\tau}
\end{array}\right)
\end{equation}
with the $2\times2$ diagonal matrix $\tilde{\tau}=\mathrm{diag}\left(\cos\alpha_{s},\,\sin\alpha_{s}\right)$.
Since the upper (lower) diagonal block is related to the $X$ ($Y$)
pocket, we have:

\begin{align*}
\Delta_{X} & =\Delta_{s}+\Delta_{d}\ ,\qquad\Delta_{Y}=\Delta_{s}-\Delta_{d}
\end{align*}

Substitution in the free energy yields two decoupled superconducting
systems:
\begin{equation}
F=\frac{\alpha}{2}\left(|\Delta_{X}|^{2}+|\Delta_{Y}|^{2}\right)+\frac{\beta}{2}\left(\big|\Delta_{X}\big|^{4}+\big|\Delta_{Y}\big|^{4}\right)
\end{equation}

\subsection{Non-zero spin-orbit coupling and inversion symmetry-breaking}

The main effect of the SOC (and also of the ISB) is to generate a
quadratic coupling between the two gaps $\Delta_{X}$ and $\Delta_{Y}$.
According to the Feynman diagram of Fig.~3 in the main text, these
terms generate the quadratic contribution to the free energy:
\begin{equation}
\delta F=\gamma\left(\Delta_{X}\Delta_{Y}^{*}+\Delta_{X}^{*}\Delta_{Y}\right)\label{EqnS:FreeEne}
\end{equation}

Here we illustrate the computation of $\gamma$ for the case of SOC.
The Feynman diagram gives:
\begin{align}
\gamma= & -\left(\frac{i\lambda}{2}\right)^{2}T\sum_{n,\fvec k}\mathrm{Tr}\left[\Delta_{X}\otimes i\sigma_{2}G_{X}(\omega_{n},\fvec K)\big(\tau_{+}\otimes\sigma_{1}+\tau_{-}\otimes\sigma_{2}\big)G_{Y}(\omega_{n},\fvec K)\Delta_{Y}^{*}\otimes(-i\sigma_{2})G_{Y}^{T}(-\omega_{n},-\fvec K)\right.\nonumber \\
 & \qquad\left.\big(\tau_{+}\otimes\sigma_{1}+\tau_{-}\otimes\sigma_{2}\big)^{T}G_{X}^{T}(-\omega_{n},-\fvec K)\right]\nonumber \\
= & -\frac{\lambda^{2}}{2}T\sum_{n,\fvec k}\left\{ \left(G_{Y}\Delta_{Y}^{*}G_{Y}^{T}\right)_{22}\left(G_{X}^{T}\Delta_{X}G_{X}\right)_{11}+\left(G_{Y}\Delta_{Y}^{*}G_{Y}^{T}\right)_{11}\left(G_{X}^{T}\Delta_{X}G_{X}\right)_{22}\right\}
\end{align}
where $G_{X/Y}=(i\omega_{n}-H_{X/Y}(K))^{-1}$ and $G_{X/Y}^{T}=(-i\omega_{n}-H_{X/Y}^{T}(-\fvec K))^{-1}$.
In general, the diagonal component of $G_{X}^{T}\Delta_{X}G_{X}$
could be either positive or negative. But if we only focus on the
projection along the band that crosses the Fermi level, this diagonal
component must be positive. Consider for instance the wave-function
$|u(\fvec K)\rangle$ that diagonalizes $H_{X}(\fvec K)$ and gives
the band $\epsilon_{u}(\fvec K)$ that crosses the Fermi surface.
We find:
\begin{equation}
\left(G_{X}^{T}\Delta_{X}G_{X}\right)_{ii}\approx\frac{\langle u(\fvec K)|\Delta_{X}|u(\fvec K)\rangle\left|\langle i|u(\fvec K)\rangle\right|^{2}}{(-i\omega_{n}-\epsilon_{u}(\fvec K))(i\omega_{n}-\epsilon_{u}(\fvec K))}=\frac{\left|\langle i|u(\fvec K)\rangle\right|^{2}}{\omega_{n}^{2}+\epsilon_{u}^{2}(\fvec K)}\langle u(\fvec K)|\Delta_{X}|u(\fvec K)\rangle
\end{equation}

Since:
\[
\langle u(\fvec k)|\Delta_{X}|u(\fvec k)\rangle=\Delta_{1}|u_{1}(\fvec k)|^{2}+\Delta_{2}|u_{2}(\fvec k)|^{2}
\]
is the SC gap projected onto the band that crosses the Fermi level,
we find that it is always positive, because both $\Delta_{1}$ and
$\Delta_{2}$ are positive, as shown in Fig.~\ref{Fig:SCPair}. Thus,
it follows that $\gamma<0$.

As for the quartic coefficients, we find that in the presence of SOC
or ISB they satisfy the relationship $\gamma_{1}-\gamma_{2}\lessapprox2\sqrt{\beta_{1}\beta_{2}}$.
As a result, the two gap functions can in principle coexist and break
time reversal symmetry at low temperatures.

\subsection{Large-momentum nematic fluctuations}

In the previous subsection, we investigated the effect of SOC and
ISB in lifting the $s$-wave/$d$-wave degeneracy. It is interesting
to study whether large-momentum nematic fluctuations, involving momentum
transfer $\fvec q\approx M=(\pi,\pi)$ and thus coupling the electron
pockets, give rise to a similar effect. We note that these large-momentum
fluctuations are actually associated with orbital order that breaks
translational symmetry (antiferro-orbital order), instead of ferro-orbital
order. The fact that the associated antiferro-orbital order has not
been observed in bulk or thin films of FeSe suggests that these fluctuations
are much smaller than the nematic ones, and therefore can be considered
a perturbation on top of the superconducting state obtained previously.
For this reason, the main contribution of the large-momentum nematic
fluctuations $\chi_{M}$ to the superconducting free energy is captured
by the Feynman diagram in Fig.~\ref{FigS:SCLargeMNem}, yielding
the quadratic term:

\begin{figure}[htbp]
\centering \includegraphics[scale=0.7]{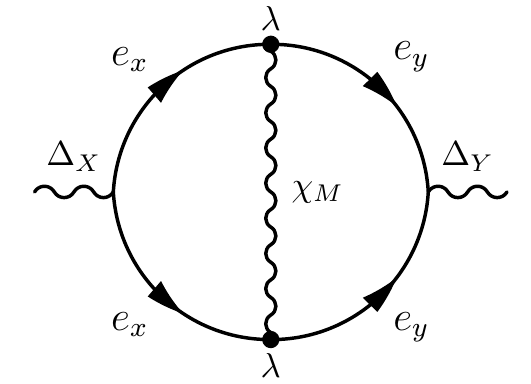} \protect\caption{Feynman diagram representing the coupling between the gaps in the
two electron pockets mediated by large-momentum nematic fluctuations
$\chi_{nem}(\fvec q\approx\fvec M)\equiv\chi_{M}$. Similar to SOC,
this coupling also lifts the degeneracy between s-wave and d-wave,
but with a much smaller effect.}

\label{FigS:SCLargeMNem}
\end{figure}

\begin{equation}
\delta F=\gamma'(\Delta_{X}^{*}\Delta_{Y}+\Delta_{Y}^{*}\Delta_{X})
\end{equation}
with
\begin{align}
\gamma'= & -T^{2}\sum_{m,n}\int\frac{\rmd^{2}\fvec k}{(2\pi)^{2}}\frac{\rmd^{2}\fvec k'}{(2\pi)^{2}}\frac{\lambda^{2}\chi(\omega_{n}-\omega_{m},\fvec k-\fvec k'+\fvec M)}{(i\omega_{n}-\epsilon_{X}(\fvec k))(-i\omega_{n}-\epsilon_{X}(-\fvec k))(i\omega_{m}-\epsilon_{Y}(\fvec k'))(-i\omega_{m}-\epsilon_{Y}(-\fvec k'))}\nonumber \\
\approx & -\lambda^{2}\chi_{M}T^{2}\sum_{m,n}\int\frac{\rmd^{2}\fvec k}{(2\pi)^{2}}\frac{\rmd^{2}\fvec k'}{(2\pi)^{2}}\frac{1}{(\omega_{n}^{2}+\epsilon_{X}^{2}(\fvec k))(\omega_{m}^{2}+\epsilon_{Y}^{2}(\fvec k'))}\nonumber \\
\approx & -N_{f}^{2}\lambda^{2}\chi_{M}\left(\ln\frac{\Lambda}{T_{c}}\right)^{2}
\end{align}
In the above formula, we made the approximation that the nematic susceptibility
is peaked at zero frequency and momentum $\mathbf{Q}$, and neglected
the orbital dependence of the coupling between the fermions and the
nematic fluctuation. It is obvious that $\gamma'<0$, showing that
the nematic fluctuation $\chi_{M}$ also favors $s$ wave.

Next, we compare the effects of large-momentum fluctuations and SOC
in lifting the degeneracy by comparing $\gamma'$ calculated here
with $\gamma$ calculated in Eq.~(\ref{EqnS:FreeEne}). From Fig.~3,
we can estimate $\gamma$ as:

\[
\gamma\approx-T\sum_{n}\int\frac{\rmd^{2}\fvec k}{(2\pi)^{2}}\frac{\lambda_{SOC}^{2}}{(\omega_{n}^{2}+\epsilon_{x}^{2}(\fvec k))(\omega_{n}^{2}+\epsilon_{y}^{2}(\fvec k))}\approx-\lambda_{SOC}^{2}T\sum_{n}\int\frac{\rmd^{2}\fvec k}{(2\pi)^{2}}\frac{1}{(\omega_{n}^{2}+\epsilon^{2}(\fvec k))^{2}}\sim-N_{f}\frac{\lambda_{SOC}^{2}}{T_{c}^{2}}
\]

Furthermore, we can use for $T_{c}$:

\begin{equation}
\ln\frac{\Lambda}{T_{c}}\approx\frac{1}{N_{f}\lambda^{2}\chi_{0}}
\end{equation}
where $\chi_{0}$ is the zero-momentum nematic susceptibility. Substituting
in the equations above, we find:
\begin{equation}
\frac{\gamma}{\gamma'}=\frac{\chi_{0}}{\chi_{M}}\frac{\lambda_{SOC}^{2}}{T_{c}^{2}}\left(\ln\frac{\Lambda}{T_{c}}\right)^{-1}\approx\left(1+2\pi^{2}\xi_{\mathrm{nem}}^{2}a^{-2}\right)\frac{\lambda_{SOC}^{2}}{T_{c}^{2}}\left(\ln\frac{\Lambda}{T_{c}}\right)^{-1}
\end{equation}
where, in the last step, we considered the expansion $\chi_{\mathrm{nem}}^{-1}(\mathbf{q})=\xi_{\mathrm{nem}}^{-2}+q^{2}$
and substituted $\chi_{0}=\chi_{\mathrm{nem}}(0)$, $\chi_{0}=\chi_{\mathrm{nem}}(\pi,\pi)$.
We now substitute reasonable, experimentally-based values for these
quantities. According to ARPES data in bulk FeSe~\cite{Zhigadlo16S},
$\lambda_{SOC}\approx10$ meV, which is of the same order as $T_{c}\approx6$
meV of the thin films. The bandwidth can be estimated as $\Lambda\sim100$
meV, whereas the nematic correlation length is certainly a few lattice
constants -- say $\xi=5a$. Substituting these numbers, we estimate
$\gamma/\gamma'\approx500\gg1$. Therefore, the effect of large-momentum
fluctuations is negligible compared to the effect of spin-orbit coupling.
A similar analysis for the inversion symmetry-breaking contribution
reveals the latter is also much larger than the large-momentum fluctuations
contribution, as long as $\lambda_{ISB}>1$ meV. Although this is
a reasonable value, first principle calculations are necessary to
estimate $\lambda_{ISB}$, which is beyond the scope of this paper.

\section{Fermi-surface projected Model}

In the previous analyses, we considered the low-energy model derived
directly from the $5\times5$ tight-binding Hamiltonian. To gain more
insight into the problem, we can further restrict our analysis only
to the bands that cross the Fermi level, since they give the dominant
contribution to the pairing instability. It is then convenient to
write the $X$ and $Y$ pockets dispersions (here $\fvec k$ refers
to the 1-Fe unit cell):
\begin{equation}
\varepsilon_{X,\mathbf{k}}=\frac{k^{2}}{2m}-\epsilon_{\mu}-\epsilon_{m}\cos2\theta\ ,\varepsilon_{Y,\mathbf{k}}=\frac{k^{2}}{2m}-\epsilon_{\mu}+\epsilon_{m}\cos2\theta
\end{equation}
and the non-interacting Hamiltonian in terms of the band operators
$f_{X}$ and $f_{Y}$:
\begin{equation}
H_{0}=\sum_{\fvec k}\varepsilon_{X,\fvec k}f_{X,\mathbf{k}}^{\dagger}f_{X,\mathbf{k}}^{\phantom{\dagger}}+\sum_{\fvec k}\varepsilon_{Y,\fvec k}f_{Y,\mathbf{k}}^{\dagger}f_{Y,\mathbf{k}}^{\phantom{\dagger}}
\end{equation}

Here, $\epsilon_{m}$ gives the mismatch between the two electron
pockets. The advantage of this model over the previous one is that
it allows us to easily tune the ratio between the mismatch and the
SOC, $\epsilon_{m}/\lambda_{\mathrm{SOC}}$, which in the full model
above is fixed by the parameter $v$ in Eq.~\ref{EqnS:Band}. To
proceed, we write down the relationship between the band operators
$f_{X}$, $f'_{X}$ and the orbital operators $d_{xy},\, d_{yz}$~\cite{Fernandes14}:
\begin{equation}
f_{X}\left(\theta\right)=\frac{\alpha d_{xy}+i\sin\theta d_{yz}}{\sqrt{\alpha^{2}+\sin^{2}\theta}}\ ;\quad f'_{X}\left(\theta\right)=\frac{i\sin\theta d_{xy}+\alpha d_{yz}}{\sqrt{\alpha^{2}+\sin^{2}\theta}}\ .
\end{equation}
where $\theta$ is the polar angle. The factor $i$ is inserted to
keep the wave-function time-reversal invariant. While $f_{X}$ describes
the band obtained from the diagonalization of $\hat{H}_{X}$ that
crosses the Fermi level, $f'_{X}$ describes the band that do not
cross the Fermi level. These relationships can be inverted to give:
\begin{equation}
d_{xy}=\frac{\alpha f_{X}-i\sin\theta f'_{X}}{\sqrt{\alpha^{2}+\sin^{2}\theta}}\ ,\qquad d_{yz}=\frac{-i\sin\theta f_{X}+\alpha f'_{X}}{\sqrt{\alpha^{2}+\sin^{2}\theta}}\ .
\end{equation}

The band operators $f_{Y}$, $f'_{Y}$, related to the $Y$ pocket,
are obtained by a rotation of $\pi/2$ followed by a mirror reflection
$\sigma_{z}$ with respect to the $(x,y)$ plane:
\[
\theta\rightarrow\theta-\frac{\pi}{2}\ ,\quad d_{xy}\rightarrow-d_{xy}\ ,\quad d_{yz}\rightarrow d_{xz}\quad\Longrightarrow\quad f_{Y}\left(\theta\right)=-\frac{\alpha d_{xy}+i\cos\theta d_{xz}}{\sqrt{\alpha^{2}+\cos^{2}\theta}}\ ;\quad f'_{Y}\left(\theta\right)=\frac{i\cos\theta d_{xy}+\alpha d_{xz}}{\sqrt{\alpha^{2}+\cos^{2}\theta}}\ .
\]

The coupling between nematic fluctuations and the band operators associated
with the $X$ electron pocket can be obtained from:
\begin{align}
H_{\mathrm{int},X} & =\lambda\sum_{\mathbf{k},\mathbf{k}'}\phi_{\mathbf{k}-\mathbf{k}'}\left(\cos\chi d_{yz,\mathbf{k}'}^{\dagger}d_{yz,\mathbf{k}}+\sin\chi d_{xy,\mathbf{k}'}^{\dagger}d_{xy,\mathbf{k}}\right)\\
H_{\mathrm{int},X} & =\lambda\sum_{\mathbf{k},\mathbf{k}'}\phi_{\mathbf{k}-\mathbf{k}'}\left(\frac{\alpha^{2}\sin\chi+\cos\chi\sin\theta\sin\theta'}{\sqrt{\big(\alpha^{2}+\sin^{2}\theta'\big)\big(\alpha^{2}+\sin^{2}\theta\big)}}\right)f_{X}^{\dagger}(\theta')f_{X}(\theta)
\end{align}
where $\chi=\tan^{-1}\big(\lambda_{2}/\lambda_{1}\big)$ and $\lambda=\sqrt{\lambda_{1}^{2}+\lambda_{2}^{2}}$
are related to the two nematic couplings $\lambda_{1}$ and $\lambda_{2}$.
Since we are interested on the states at the Fermi level, we hereafter
focus only on the contributions arising from bilinear combinations
of $f_{X}$ and $f_{Y}$, since the band dispersions corresponding
to $f'_{X}$ and $f'_{Y}$ do not cross the Fermi surface. Similarly,
for pocket $Y$ we find:
\begin{equation}
H_{\mathrm{int},Y}=-\lambda\sum_{\mathbf{k},\mathbf{k}'}\phi_{\mathbf{k}-\mathbf{k}'}\left(\frac{\alpha^{2}\sin\chi+\cos\chi\cos\theta\cos\theta'}{\sqrt{\big(\alpha^{2}+\cos^{2}\theta'\big)\big(\alpha^{2}+\cos^{2}\theta\big)}}\right)f_{Y}^{\dagger}(\theta')f_{Y}(\theta)
\end{equation}

The linearized gap equations for the two pockets then become:
\begin{align}
\eta\Delta_{X}(\theta) & =\lambda^{2}\chi_{\mathrm{nem}}N_{0}\ln\frac{\Lambda}{T}\int\frac{\rmd\theta'}{2\pi}\frac{\left(\alpha^{2}\sin\chi+\cos\chi\sin\theta\sin\theta'\right)^{2}}{\big(\alpha^{2}+\sin^{2}\theta'\big)\big(\alpha^{2}+\sin^{2}\theta\big)}\Delta_{X}(\theta')\ ,\\
\eta\Delta_{Y}(\theta) & =\lambda^{2}\chi_{\mathrm{nem}}N_{0}\ln\frac{\Lambda}{T}\int\frac{\rmd\theta'}{2\pi}\frac{\left(\alpha^{2}\sin\chi+\cos\chi\cos\theta\cos\theta'\right)^{2}}{\big(\alpha^{2}+\cos^{2}\theta'\big)\big(\alpha^{2}+\cos^{2}\theta\big)}\Delta_{Y}(\theta')\ ,
\end{align}
where $\Lambda$ is the high energy cutoff, and $N_{0}$ is the density
of states. These gap equations can be conveniently parametrized and
solved in terms of the intra-orbital gaps $\Delta_{1}$ and $\Delta_{2}$:
\begin{align}
\Delta_{X}\left(\theta\right) & =\frac{\Delta_{1}\alpha^{2}+\Delta_{2}\sin^{2}\theta}{\alpha^{2}+\sin^{2}\theta}\\
\Delta_{Y}\left(\theta\right) & =\frac{\Delta_{1}\alpha^{2}+\Delta_{2}\cos^{2}\theta}{\alpha^{2}+\cos^{2}\theta}
\end{align}

\begin{figure}[htbp]
\centering \subfigure[\label{FigS:ModelSCSSOC:Eig}]{\includegraphics[scale=0.5]{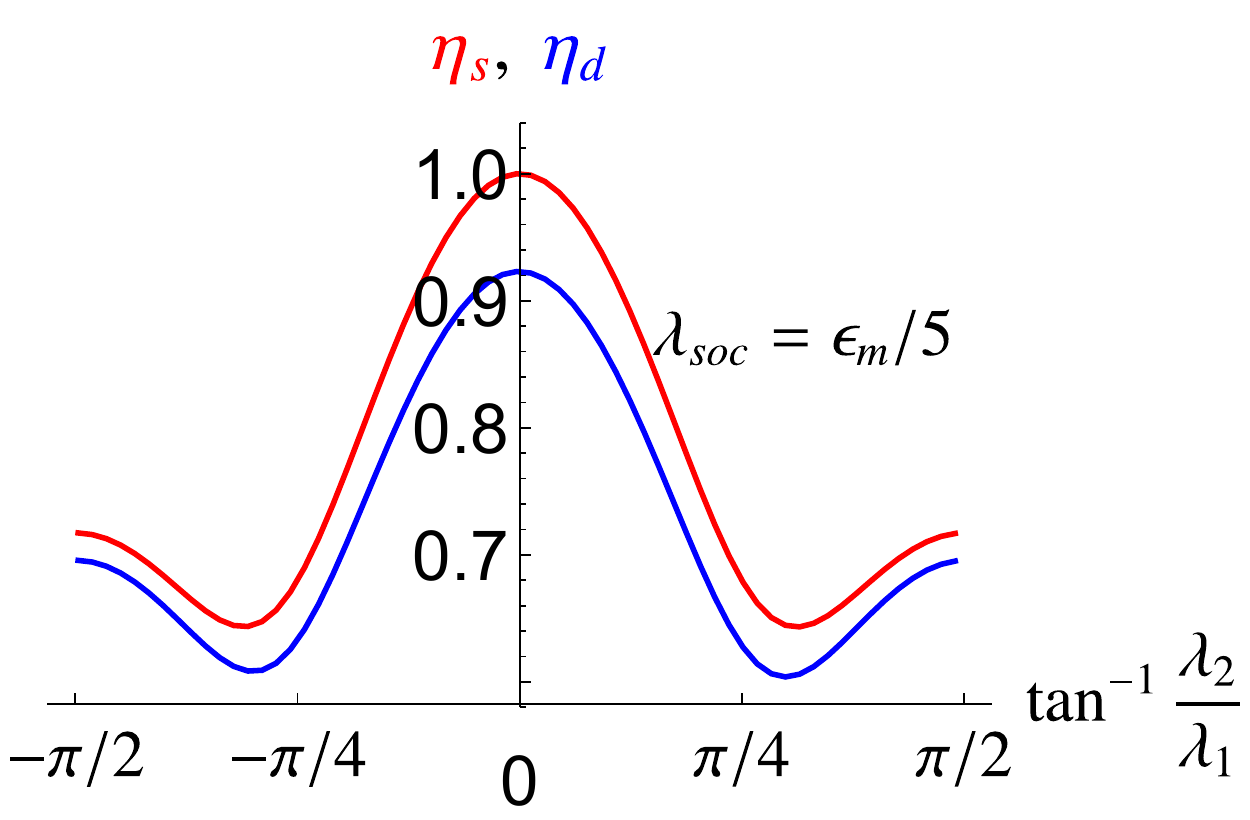}}
\subfigure[\label{FigS:ModelSCSSOC:Coupling1}]{\includegraphics[scale=0.4]{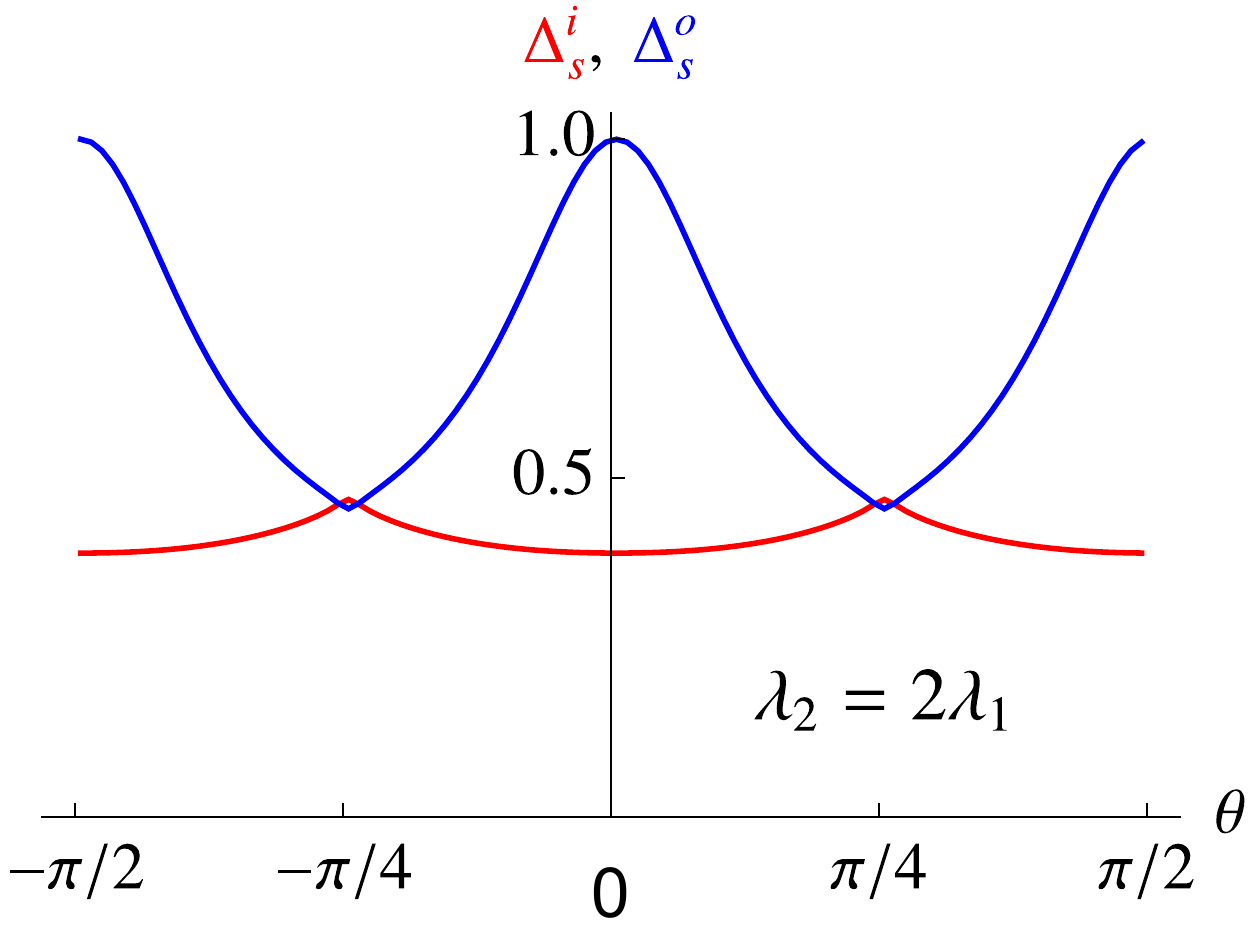}}
\subfigure[\label{FigS:ModelSCSSOC:Coupling2}]{\includegraphics[scale=0.4]{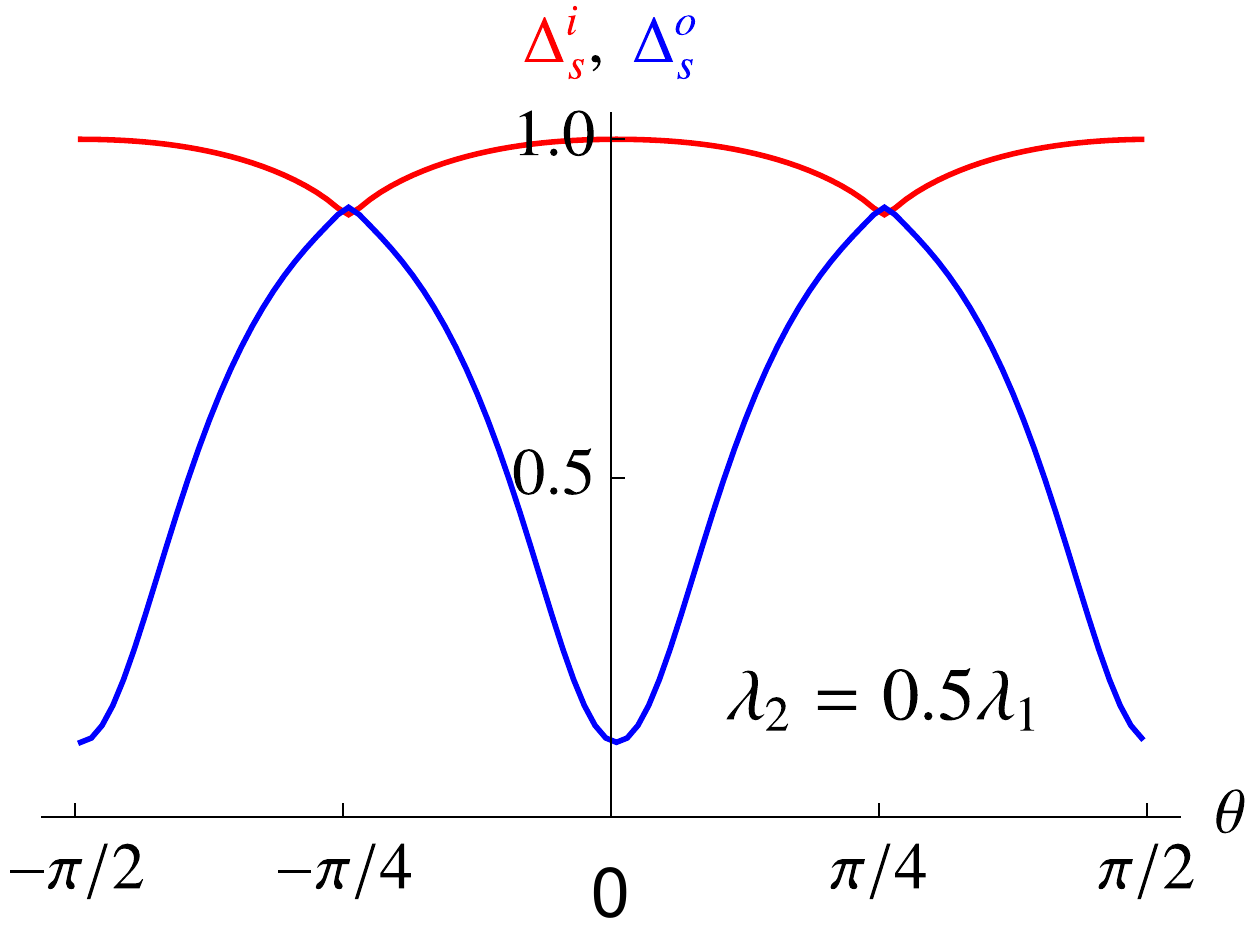}}
\protect\protect\protect\protect\protect\caption{SC in different channels for $\alpha=0.5$ and $\lambda_{\mathrm{SOC}}=5\epsilon_{m}$.
(a) The eigenvalues of the $A_{1g}$ ($s$-wave) and $B_{2g}$ ($d$-wave)
pairing channels. (b) and (c): Projected gap functions of the $A_{1g}$
solution onto the inner and outer electron pockets for $\lambda_{2}=2\lambda_{1}$
and $\lambda_{2}=0.5\lambda_{1}$, respectively.}

\label{FigS:ModelSCSSOC}
\end{figure}

To calculate the gap function in the presence of SOC and ISB, we need
to write down the two additional non-interacting terms in the band
basis. The SOC is given by
\begin{equation}
H_{\mathrm{SOC}}=\frac{i}{2}\lambda_{\mathrm{SOC}}\sum_{\fvec k}\left[d_{xz,\fvec k+\fvec Q_{Y}}^{\dagger}\sigma_{1}d_{xy,\fvec k+\fvec Q_{X}}+d_{xy,\fvec k+\fvec Q_{Y}}^{\dag}\sigma_{2}d_{yz,\fvec k+\fvec Q_{X}}\right]+h.c.
\end{equation}

Projecting onto the Fermi surface, we find:
\begin{equation}
H_{\mathrm{SOC}}=\lambda_{\mathrm{SOC}}\sum_{\fvec k}\frac{\alpha}{2\sqrt{\alpha^{2}+\cos^{2}\theta}\sqrt{\alpha^{2}+\sin^{2}\theta}}f_{Y}^{\dag}\left(\cos\theta\sigma_{1}-\sin\theta\sigma_{2}\right)f_{X}+h.c.
\end{equation}

Similarly, the ISB term is:
\begin{equation}
H_{\mathrm{ISB}}=\lambda_{\mathrm{ISB}}\sum_{\fvec k}d_{xz,\fvec k+\fvec Q_{Y}}^{\dagger}d_{yz,\fvec k+\fvec Q_{X}}+h.c.
\end{equation}
whose projection onto the Fermi surface gives:
\begin{equation}
H_{\mathrm{ISB}}=-\lambda_{\mathrm{ISB}}\sum_{\fvec k}\frac{\sin\theta\cos\theta}{\sqrt{\alpha^{2}+\cos^{2}\theta}\sqrt{\alpha^{2}+\sin^{2}\theta}}\, f_{X}^{\dagger}f_{Y}+h.c.\
\end{equation}

\begin{figure}[htbp]
\centering \includegraphics[scale=0.5]{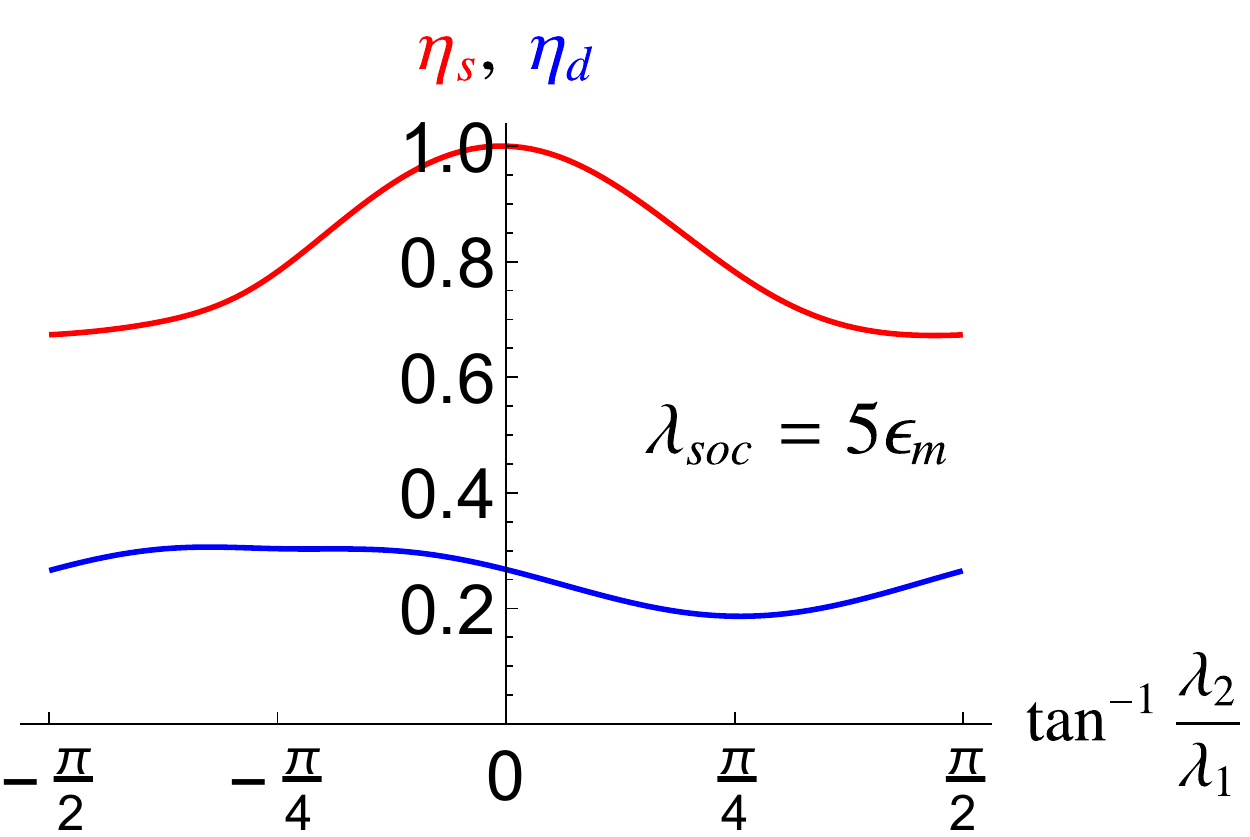} \protect\protect\protect\protect\protect\caption{Eigenvalues of the $A_{1g}$ ($s$-wave) and $B_{2g}$ ($d$-wave)
pairing channels for $\alpha=0.5$ and $\lambda_{\mathrm{SOC}}=5\epsilon_{m}$.}

\label{FigS:ModelSCLSOC}
\end{figure}

The results for the case of SOC are shown in Fig.~\ref{FigS:ModelSCSSOC}
(for $\lambda_{\mathrm{SOC}}\ll\epsilon_{m}$) and \ref{FigS:ModelSCLSOC}
(for $\lambda_{\mathrm{SOC}}\gg\epsilon_{m}$). In the former case,
the angular dependence of the gap functions in the inner and outer
pockets, and the splitting of the $A_{1g}$ and $B_{2g}$ degeneracies,
are similar to Fig.~2 of the main text, which was obtained using
the full orbital model. In the latter case, the degeneracy lifting
is more pronounced, and the gaps in the two pockets are similar, as
shown in Fig.~4 of the main text.

\begin{figure}[htbp]
\centering \subfigure[\label{Fig:ModelSCLInv:Eigen}]{\includegraphics[scale=0.4]{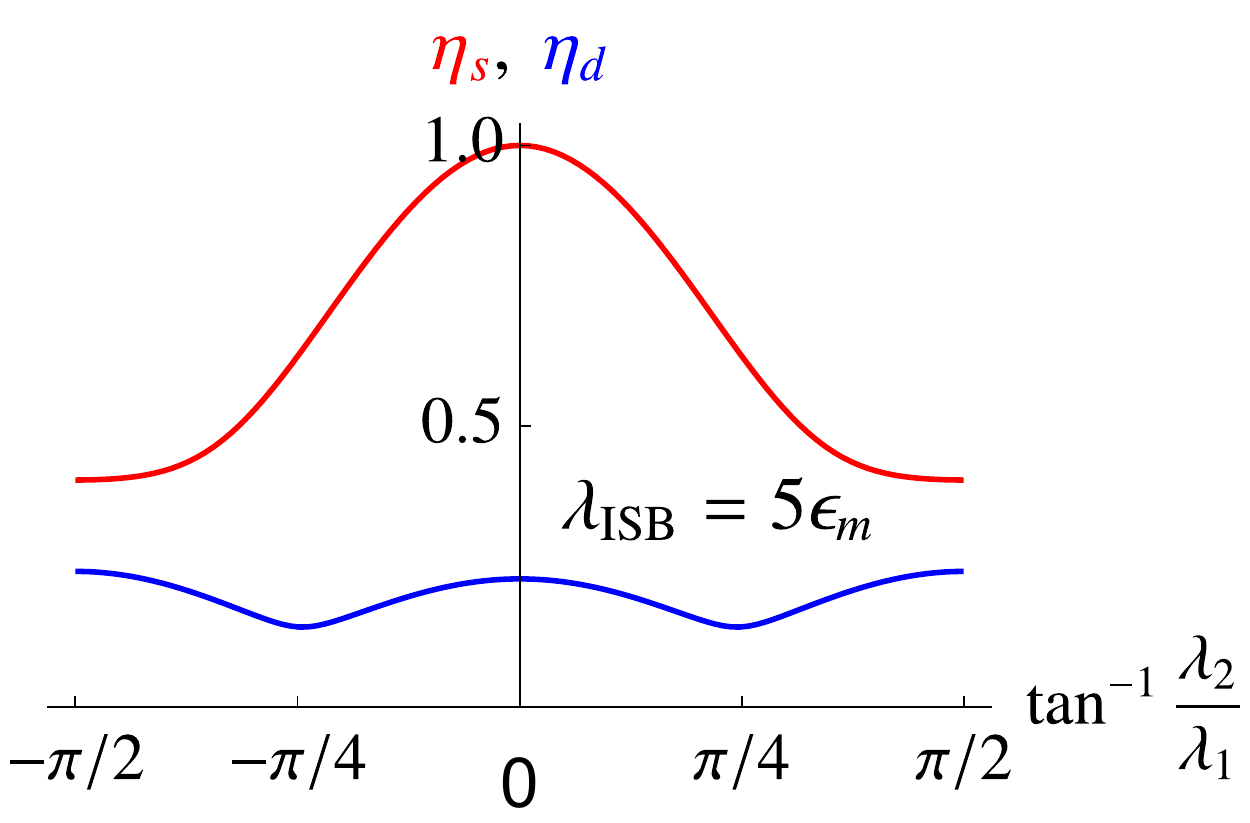}}
\subfigure[\label{Fig:ModelSCLInv:Coupling1}]{\includegraphics[scale=0.4]{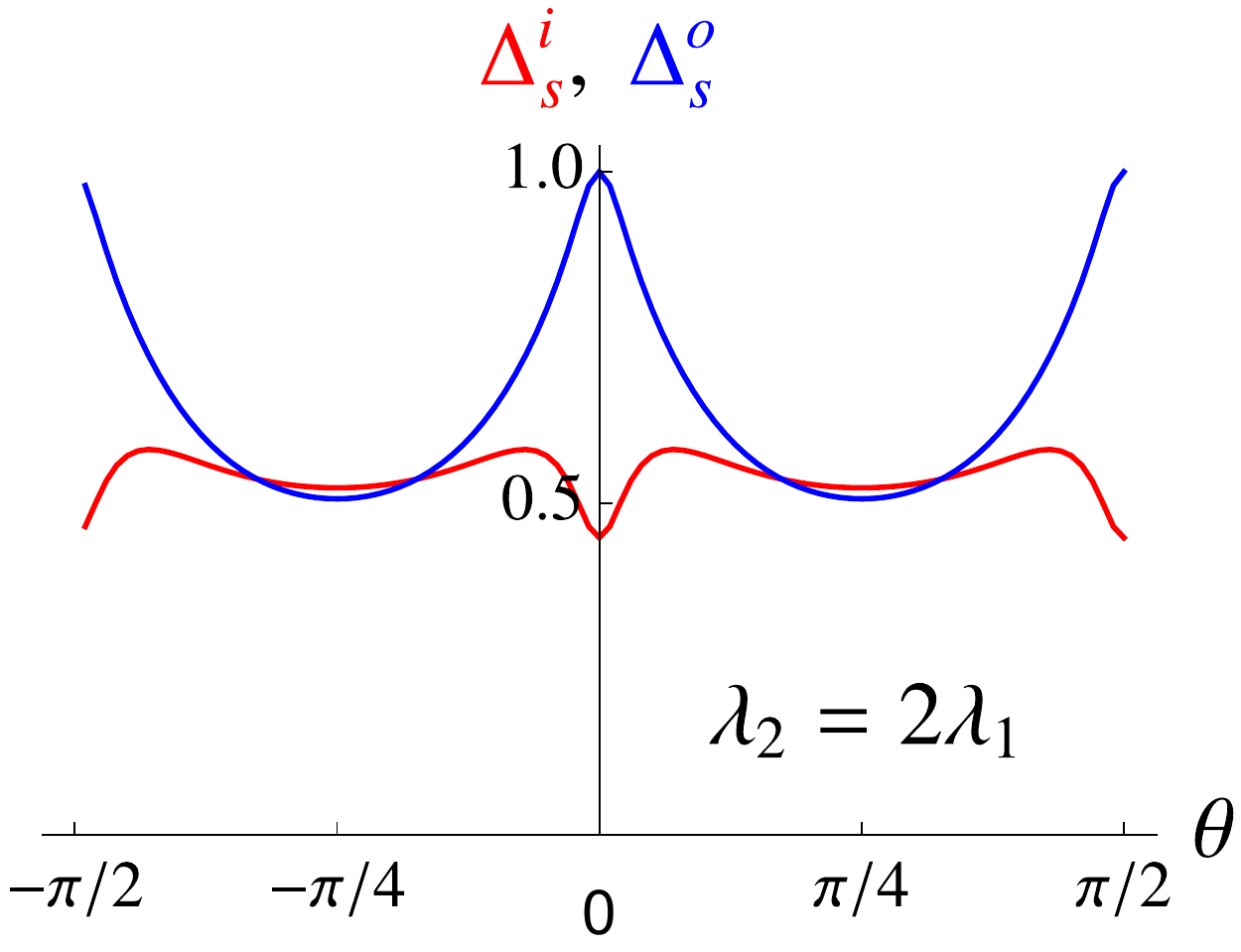}}
\subfigure[\label{Fig:ModelSCLInv:Coupling2}]{\includegraphics[scale=0.4]{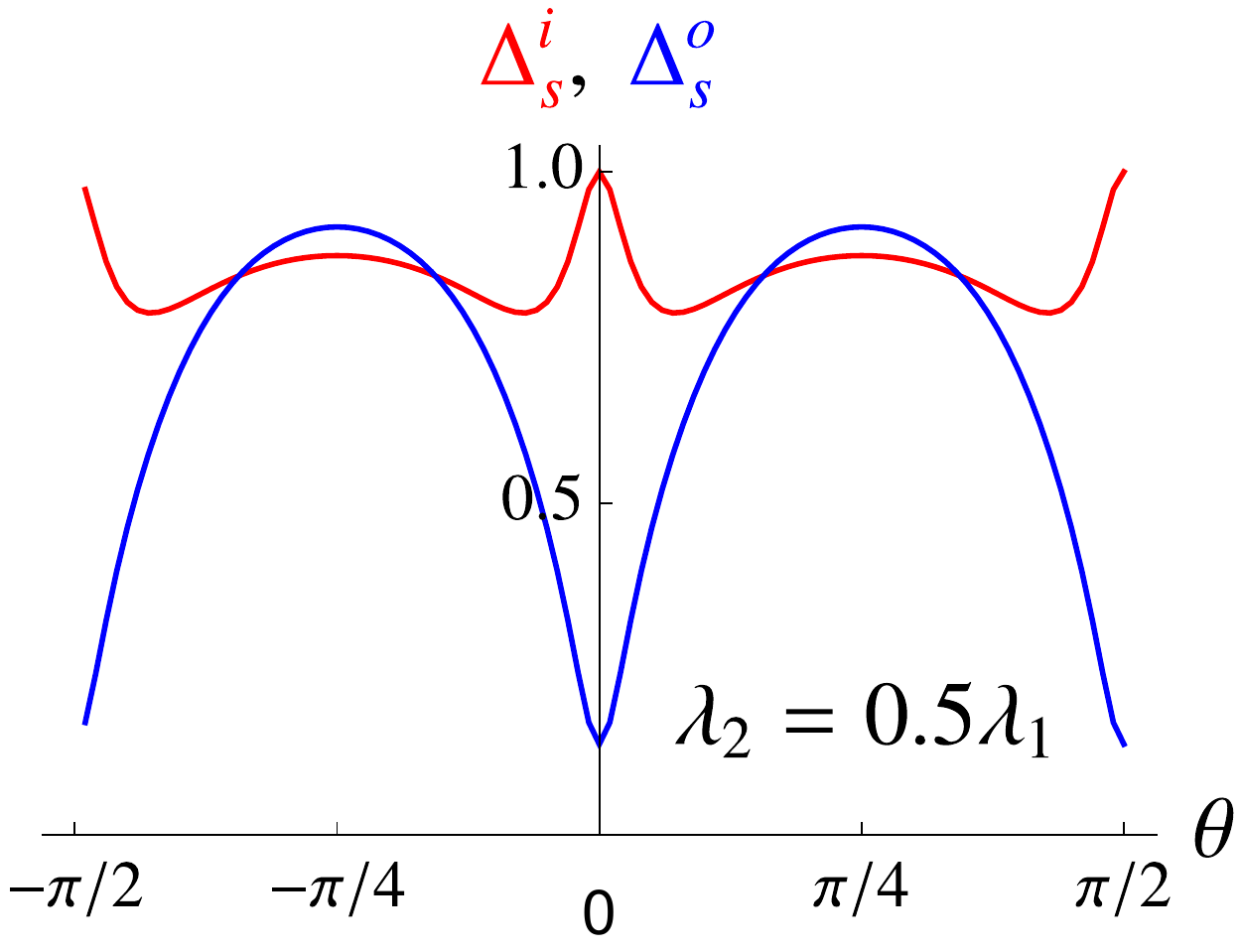}}
\protect\protect\protect\protect\protect\caption{SC in different channels for $\alpha=0.5$ and $\lambda_{\mathrm{ISB}}=5\epsilon_{m}$.
(a) The eigenvalues of the $A_{1g}+A_{2u}$ ($s$-wave) and $B_{2g}$
($d$-wave) pairing channels. (b) and (c): Projected gap functions
of the $A_{1g}+A_{2u}$ solution onto the inner and outer electron
pockets for $\lambda_{2}=2\lambda_{1}$ and $\lambda_{2}=0.5\lambda_{1}$,
respectively.}
\label{Fig:ModelSCLInv}
\end{figure}

The impact of ISB on SC is similar to the case of SOC. As shown in
Fig.~\ref{Fig:ModelSCLInv} (for $\lambda_{\mathrm{ISB}}\gg\epsilon_{m}$),
the $A_{1g}+A_{2u}$ ($s$-wave) pairing is the leading instability.
The angular dependence of the SC gap is also similar to the case of
large SOC. The sharp peaks or troughs at $\theta=0$ and $\pm\pi/2$
are a consequence of the fact that the effective ISB term $\lambda_{\mathrm{ISB}}$
vanishes at these points of the Fermi surface.

\begin{thebibliography}{10}
\bibitem{Mazin08} I. I. Mazin, D. J. Singh, M. D. Johannes, M.H.
Du, Phys. Rev. Lett. \textbf{101}, 057003 (2008).

\bibitem{Hirschfeld11} P. J. Hirschfeld, M. M. Korshunov, and I.
I. Mazin, Rep. Prog. Phys. \textbf{74}, 124508 (2011).

\bibitem{Chubukov11} D. N. Basov and A. V. Chubukov, Nature Phys.
\textbf{7}, 272 (2011).

\bibitem{Chubukov_review} A. V. Chubukov, Annu. Rev. Condens. Matter
Phys. \textbf{3}, 5792 (2012).

\bibitem{Xue12} Q. Y. Wang, \emph{et al.}, Chin. Phys. Lett. \textbf{29},
037402 (2012).

\bibitem{Feng12} D. F. Liu \emph{et al.}, Nat. Commun. \textbf{3},
931 (2012).

\bibitem{Zhou13} S. L. He \emph{et al.}, Nat. Mater. \textbf{12},
605 (2013).

\bibitem{Feng13} S. Tan \emph{et al.}, Nat. Mater. \textbf{12}, 634
(2013).

\bibitem{Xue14} W. H. Zhang, \emph{et al.}, Chin. Phys. Lett. \textbf{31},
017401 (2014).

\bibitem{Jia14} J.-F. Ge, Z.-L. Liu, C. Liu, C.-L. Gao, D. Qian,
Q.-K. Xue, Y. Liu, and J.-F. Jia, Nature Mater., \textbf{14}, 285
(2014).

\bibitem{Ding15} P. Zhang, \emph{et al.}, Phys. Rev. B \textbf{94},
104510 (2016).

\bibitem{FengNC14} R. Peng, \emph{et al}., Nature Commun. \textbf{5},
5044 (2014).

\bibitem{Chen16} B. Lei, \emph{et al}., Phys. Rev. Lett. \textbf{116},
077002 (2016).

\bibitem{Wang14} Y. Sun, W. Zhang, Y. Xing, F. Li, Y. Zhao, Z. Xia,
L. Wang, X. Ma, Q.-K. Xue, and J. Wang, Sci. Rep. \textbf{4}, 6040
(2014).

\bibitem{Khasanov09} E. Pomjakushina, K. Conder, V. Pomjakushin,
M. Bendele, and R. Khasanov, Phys. Rev. B \textbf{80}, 024517 (2009).

\bibitem{Shen14} J. J. Lee, \emph{et al.}, Nature \textbf{515}, 245
(2014).

\bibitem{HSU08} F.-C. Hsu, \emph{et al.}, Proc. Natl Acad. Sci. \textbf{105}
14262 (2008).

\bibitem{Lee12} Y.-Y. Xiang, F. Wang, D. Wang, Q.-H. Wang, and D.-H.
Lee, Phys. Rev. B \textbf{86}, 134508 (2012).

\bibitem{Zhao16} Y. C. Tian, W. H. Zhang, F. S. Li, Y. L. Wu, Q.
Wu, F. Sun, G. Y. Zhou, L. Wang, X. Ma, Q.-K. Xue, and J. Zhao, Phys.
Rev. Lett. \textbf{116}, 107001 (2016).

\bibitem{Rademaker16} L. Rademaker, Y. Wang, T. Berlijn, and S. Johnston,
New J. Phys. 18, 022001 (2016).

\bibitem{Millis16} Y. Zhou and A. J. Millis, Phys. Rev. B \textbf{93},
224506 (2016).

\bibitem{DHLee_STO} Z.-X. Li, F. Wang, H. Yao, and D. H. Lee, Science
Bulletin \textbf{61}, 925 (2016).

\bibitem{Johnston16} Y. Wang, K. Nakatsukasa, L. Rademaker, T. Berlijn,
and S. Johnston, Supercond. Sci. Technol. \textbf{29}, 054009 (2016).

\bibitem{Dolgov16} M. L. Kulic and O. V. Dolgov, arXiv:1607.00843.

\bibitem{Tsukazaki16} J. Shiogai, Y. Ito, T. Mitsuhashi, T. Nojima,
and A. Tsukazaki, Nature Phys. \textbf{12} 42 (2016).

\bibitem{ShenSC15} Z. R. Ye, \emph{et al}., arXiv:1512.02526.

\bibitem{Takahashi15} Y. Miyata, K. Nakayama, K. Sugawara, T. Sato,
and T. Takahashi, Nature Mater. \textbf{14}, 775 (2015).

\bibitem{FengPRB15} X. H. Niu, \emph{et al}., Phys. Rev. B 92, 060504
(2015).

\bibitem{Zhou16} L. Zhao, \emph{et al}., Nat. Commun. \textbf{7},
10608 (2016).

\bibitem{Coldea15} M. D. Watson, \emph{et al}., Phys. Rev. B \textbf{91},
155106 (2015).

\bibitem{RMF15} A. V. Chubukov, R. M. Fernandes, and J. Schmalian,
Phys. Rev. B \textbf{91}, 201105 (2015).

\bibitem{RMF16} A. V. Chubukov, M. Khodas, and R. M. Fernandes, arXiv:1602.05503.

\bibitem{DHLee_FeSe15} F. Wang, S. A. Kivelson, and D.-H. Lee, Nature
Phys. \textbf{11}, 959 (2015).

\bibitem{Si15} R. Yu and Q. Si, Phys. Rev. Lett. \textbf{115}, 116401
(2015).

\bibitem{Glasbrenner15} J. K. Glasbrenner, I. I. Mazin, H. O. Jeschke,
P. J. Hirschfeld, R. M. Fernandes, and R. Valenti, Nature Phys. \textbf{11},
953 (2015).

\bibitem{ShenNem15} Y. Zhang, \emph{et al}., Phys. Rev. B \textbf{94},
115153 (2016).

\bibitem{Feng16} C. H. P. Wen, \emph{et al}., Nat. Commun. \textbf{7}
10840 (2016).

\bibitem{Hoffman16} D. Huang, T. A. Webb, S. Fang, C.-L. Song, C.-Z.
Chang, J. S. Moodera, E. Kaxiras, and J. E. Hoffman, Phys. Rev. B
\textbf{93}, 125129 (2016).

\bibitem{Yamase13} H. Yamase and R. Zeyher, Phys. Rev. B \textbf{88},
180502(R) (2013).

\bibitem{Vishwanath15} P. T. Dumitrescu, M. Serbyn, R. T. Scalettar,
and A. Vishwanath, Phys. Rev. B \textbf{94}, 155127 (2016).

\bibitem{Zhigadlo16} S. V. Borisenko, \emph{et al}., Nat. Phys. \textbf{12}
311 (2016).

\bibitem{FengNP15} Q. Fan \emph{et al.}, Nature Phys. \textbf{11},
946 (2015).

\bibitem{FengPRL14} R. Peng, \emph{et al}., Phys. Rev. Lett. \textbf{112},
107001 (2014).

\bibitem{Wen16} Z. Du, X. Yang, H. Lin, D. Fang, G. Du, J. Xing,
H. Yang, X. Zhu, H.-H. Wen, Nat. Commun. \textbf{7}, 10565(2016).

\bibitem{Oskar13} V. Cvetkovic and O. Vafek, Phys. Rev. B \textbf{88},
134510 (2013).

\bibitem{Vafek14} R. M. Fernandes and O. Vafek, Phys. Rev. B \textbf{90},
214514 (2014).

\bibitem{Schattner15} Y. Schattner, S. Lederer, S. A. Kivelson, and
E. Berg, Phys. Rev. X \textbf{6}, 031028 (2016).

\bibitem{DHLee15} Z-X.~Li, F.~Wang, H.~Yao, and D.~H.~Lee, arXiv:1512.04541.

\bibitem{note} To be consistent with previous works, the irreducible
representations refer to the actual crystallographic 2-Fe Brillouin
zone coordinate system $(K_{x},K_{y})$, i.e. $B_{2g}$ transforms
as $k_{x}^{2}-k_{y}^{2}$ (or $K_{x}K_{Y}$) and $B_{1g}$ transforms
as $k_{x}k_{y}$ (or $K_{x}^{2}-K_{y}^{2}$).

\bibitem{Borisenko} A. Fedorov, \emph{et al.}, arXiv:1606.03022.

\bibitem{Fernandes13} R. M. Fernandes and A. J. Millis, Phys. Rev.
Lett. \textbf{110}, 117004 (2013).

\bibitem{DHLee_13} F. Yang, F. Wang, and D.-H. Lee, Phys. Rev. B
\textbf{88}, 100504 (2013).

\bibitem{Brydon14} P. M. R. Brydon, S. Das Sarma, Hoi-Yin Hui, Jay
D. Sau, Phys. Rev. B \textbf{90}, 184512 (2014).

\bibitem{YXWang16} Y. Wang, G. Y. Cho, T. L. Hughes, and E. Fradkin,
Phys. Rev. B \textbf{93}, 134512 (2016).

\bibitem{Kivelson15} S. Lederer, Y. Schattner, E. Berg, and S. A.
Kivelson, Phys. Rev. Lett. \textbf{114}, 097001 (2015).

\bibitem{Zhang15} Y. Zhang, J. J. Lee, R. G. Moore, W. Li, M. Yi,
M. Hashimoto, D. H. Lu, T. P. Devereaux, D.-H. Lee, and Z.-X. Shen,
Phys. Rev. Lett. \textbf{117}, 117001 (2016).

\bibitem{Hu14} N. Hao and J. Hu, Phys. Rev. X \textbf{4}, 031053
(2014).

\bibitem{Khodas12} M. Khodas and A. V. Chubukov, Phys. Rev. Lett.
\textbf{108}, 247003 (2012).

\bibitem{Hinojosa15} A. Hinojosa and A. V. Chubukov, Phys. Rev. B
\textbf{91}, 224502 (2015).

\bibitem{Xing14} B. Li, Z. W. Xing, G. Q. Huang, and D. Y. Xing,
J. Appl. Phys. \textbf{115}, 193907 (2014)¡£



\end{thebibliography}

\begin{thebibliography}{1}
\bibitem{Oskar} V. Cvetkovic and O. Vafek, Phys. Rev. B \textbf{88},
134510 (2013).

\bibitem{Fernandes14} J. Kang, A. F. Kemper, and R. M. Fernandes,
Phys. Rev. Lett. \textbf{113}, 217001 (2014).

\bibitem{Zhigadlo16S} S. V. Borisenko, \emph{et al}., Nat. Phys. \textbf{12}
311 (2016). 
\end{thebibliography}
\end{document}